\pdfoutput=1
\documentclass[times]{speauth}

\usepackage{multirow}
\usepackage{pbox}
\usepackage{hyperref}
\usepackage[norefs]{refcheck}
\usepackage{placeins}
\newcommand{\microop}{$\mu$op}
\newcommand{\microops}{\microop{}s}
\usepackage{underscore}
\usepackage{threeparttable}
\usepackage{colortbl}
\usepackage{times}
\usepackage{subfig}
\usepackage{alphalph}
\usepackage{float}
\usepackage{listings}
\usepackage{balance}
\lstdefinestyle{customc}{%
  belowcaptionskip=1\baselineskip,
  breaklines=true,
  xleftmargin=\parindent,
  language=C,
  showstringspaces=false,
  basicstyle=\small\ttfamily,
  keywordstyle=\bfseries\color{green!40!black},
  numberstyle=\tiny,
  stepnumber=2, numbersep=5pt
  commentstyle=\itshape\color{purple!40!black},
  identifierstyle=\bfseries\color{black},
  stringstyle=\color{orange},
   morekeywords={uint64_t,uint32_t,__m256i,__m128i,UINT64_C},
}

\usepackage{arydshln}
\cgsn{Natural Sciences and Engineering Research Council of Canada}{261437}
\newcommand{\Censeighten}{\textsc{Census1881}}
\newcommand{\CensInc}{\textsc{CensusInc}}
\newcommand{\Wikileaks}{\textsc{Wikileaks}}
\newcommand{\Weather}{\textsc{Weather}}

\newcommand{\Censeightensrt}{\textsc{Census1881}$^{\mathrm{sort}}$}
\newcommand{\CensIncsrt}{\textsc{CensusInc}$^{\mathrm{sort}}$}
\newcommand{\Wikileakssrt}{\textsc{Wikileaks}$^{\mathrm{sort}}$}
\newcommand{\Weathersrt}{\textsc{Weather}$^{\mathrm{sort}}$}

\usepackage{listings}
\usepackage{booktabs}

\usepackage{microtype}
\usepackage{inconsolata}
\usepackage{tikz}
\usetikzlibrary{chains,arrows,automata,decorations.markings,positioning,calc,decorations.pathreplacing,patterns}

\lstdefinestyle{customc}{%
  belowcaptionskip=1\baselineskip,
  breaklines=true,
  xleftmargin=\parindent,
  language=C,
  showstringspaces=false,
  basicstyle=\small\ttfamily,
  keywordstyle=\bfseries\color{green!40!black},
  numberstyle=\tiny,
  stepnumber=2, numbersep=5pt
  commentstyle=\itshape\color{purple!40!black},
  identifierstyle=\bfseries\color{black},
  stringstyle=\color{orange},
   morekeywords={uint64_t,uint32_t,__m256i,__m128i,UINT64_C},
}

\usepackage[english]{babel}
\usepackage[utf8x]{inputenc}
\usepackage{amsmath}
\usepackage{url}
\usepackage{underscore}
\usepackage{siunitx}
\usepackage[colorinlistoftodos]{todonotes}
\usepackage{algorithm}
\usepackage[noend]{algorithmic}
\usepackage{pdfcomment}

\usepackage{graphicx}

\newcommand\ourtitle{Roaring Bitmaps: Implementation of an Optimized Software Library}

\makeatletter
\def\ps@pprintTitle{%
  \let\@oddhead\@empty
  \let\@evenhead\@empty
  \let\@oddfoot\@empty
  \let\@evenfoot\@oddfoot
}

\makeatother

\begin{document}

\title{\ourtitle{}}



\runningheads{D.~Lemire \textit{et al.}}{\ourtitle{}}

\author{Daniel~Lemire\affil{1},  Owen~Kaser\affil{2}, Nathan~Kurz\affil{3}, Luca~Deri\affil{4}, Chris~O'Hara\affil{5}, François~Saint-Jacques\affil{6}, Gregory~Ssi-Yan-Kai\affil{7} }

\address{\affilnum{1}Université du Québec (TELUQ), Montreal, QC, Canada\break
\affilnum{2}Computer Science,
UNB Saint John, Saint John, NB, Canada\break
\affilnum{3}Orinda, California USA\break
\affilnum{4}IIT/CNR, ntop,
Pisa, Italy\break
\affilnum{5}Kissmetrics,
Bay Area, California, USA\break
\affilnum{6}AdGear Technologies, Inc.,
Montreal, QC, Canada\break
\affilnum{7}42 Quai Georges Gorse,
Boulogne-Billancourt, France
}

\cgsn{Natural Sciences and Engineering Research Council of Canada}{261437}
\corraddr{Daniel Lemire, TELUQ,
Universit\'e du Qu\'ebec,
5800 Saint-Denis,
Office 1105,
Montreal (Quebec),
H2S 3L5 Canada. Email: lemire@gmail.com}


\begin{abstract}Compressed bitmap indexes
are used in systems
such as Git or Oracle to accelerate
queries.
They  represent sets and often support operations such as  unions, intersections, differences, and symmetric differences.
Several important systems such as Elasticsearch, Apache Spark, Netflix's Atlas, LinkedIn's Pinot, Metamarkets' Druid, Pilosa, Apache Hive, Apache Tez, Microsoft Visual Studio Team Services and Apache Kylin rely on a specific type of compressed bitmap index called Roaring.
We present an optimized software library written in C implementing Roaring bitmaps: CRoaring. It benefits from several algorithms designed for the single-instruction-multiple-data (SIMD) instructions available on commodity processors. In particular, we present vectorized algorithms to compute the intersection, union, difference and symmetric difference between arrays. We benchmark the library against a wide range of competitive alternatives, identifying weaknesses and strengths in our software. Our work is
available under a liberal open-source license.
\end{abstract}


\maketitle

\lstset{escapechar=@,style=customc}

\section{Introduction}

Contemporary computing hardware offers performance opportunities through improved  parallelism, by having more cores and better  single-instruction-multiple-data (SIMD) instructions.
Meanwhile, software indexes often determine the performance of big-data applications. Efficient indexes not only improve latency and throughput, but they also reduce energy usage~\cite{graefe2008database}.

Indexes are often made of sets of numerical identifiers (stored as integers).
For instance,
inverted indexes map query terms to document identifiers in search engines,
and conventional database indexes map column values to record identifiers.
We often need efficient computation of the intersection ($A\cap B$), the union  ($A\cup B$), the difference ($A	\setminus B$), or the symmetric difference ($(A	\setminus B) \cup (B	\setminus A)$)
of these sets.


The bitmap (or bitset) is a time-honored strategy to represent sets of integers concisely. Given a universe of $n$~possible integers, we use a vector of $n$~bits to represent any one set. On a 64-bit processor, $\lceil n /64 \rceil$~inexpensive
bitwise 
operations suffice to compute set operations between two bitmaps:
\begin{itemize}
\item the intersection corresponds to the bitwise AND;
\item the union corresponds to the bitwise OR;
\item the difference corresponds to the bitwise ANDNOT;
\item the symmetric difference corresponds to the bitwise XOR.
\end{itemize}
Unfortunately, when the range of possible values ($n$) is too wide, bitmaps can be too large to be practical. For example, it might be impractical to represent the set $\{1, 2^{31}\}$ using a bitset. For this reason, we often  use compressed bitmaps.

Though there are many ways to compress bitmaps (see \S~\ref{sec:compressedbitsets}),
several systems rely on an approach called Roaring including
 Elasticsearch~\cite{RoaringDocIdSetBlogPost}, Metamarkets' Druid~\cite{Chambi:2016:ODR:2938503.2938515},
eBay's Apache Kylin~\cite{kylin}, Netflix's Atlas~\cite{atlas}, LinkedIn's Pinot~\cite{pinot}, Pilosa~\cite{pilosa}, Apache Hive, Apache Tez, Microsoft Visual Studio Team Services~\cite{MVSTS}, and Apache Spark~\cite{Zaharia:2010:SCC:1863103.1863113,Interlandi:2015:TDP:2850583.2850595}. In turn, these systems are in widespread use: e.g.,  Elasticsearch provides the search capabilities of Wikipedia~\cite{wiki:CirrusSearch}. Additionally, Roaring is used in machine learning~\cite{abuzaid2016yggdrasil}, in data visualization~\cite{Siddiqui:2016:EDE:3025111.3025126}, in natural language processing~\cite{nlp2016}, in RDF databases~\cite{Fokou:2015:CTS:2950939.2950988}, and in geographical information systems~\cite{Krogh:2016:EII:2996913.2996972}.
Wang et al.\ recommend Roaring bitmaps as a superior alternative~\cite{goroaring2017} to other compressed bitmaps.

Roaring partitions the space $[0,2^{32})$ into \emph{chunks} consisting of ranges of  $2^{16}$~integers ($[0,2^{16}), [2^{16}, 2^{17}),  \ldots$)~\cite{SPE:SPE2325,lemire2016consistently}.\footnote{There are 64-bit extensions to Roaring~\cite{RNTI/papers/1002250}.}
For a value in the set, its least significant sixteen~bits are stored in a container corresponding to its chunk (as determined by its most significant sixteen~bits), using one of three possible container types:
\begin{itemize}\itemsep0em
\item bitset containers made of $2^{16}$~bits or \SI{8}{kB};
\item array containers made of up to \num{4096}~sorted 16-bit integers;
\item run containers made of a series of sorted $\langle s,l\rangle$ pairs indicating that all integers in the range $[s, s+l]$ are present.\footnote{The Roaring implementation~\cite{RoaringDocIdSetBlogPost} found in Elasticsearch lacks run containers but has ``negated'' array containers, where all but the missing values are stored.}
\end{itemize}
At a high level, we can view a Roaring bitmap as a list of 16-bit
numbers (corresponding to the most-significant \SI{2}{B}
of the
values present in
the set)%
,
each of which is coupled with a reference to a container
holding another set of 16-bit numbers corresponding to the
least-significant \SI{2}{B} of the elements sharing the same prefix.
See Fig.~\ref{fig:roaringbitmap}. 

We dynamically pick the container  type to minimize memory usage. For example, when intersecting two bitset containers, we determine whether the result is an array or a bitset container on-the-fly. As we add or remove values, a container's type might change. No bitset container may store fewer than \num{4097}~distinct values; no array container may store more than \num{4096}~distinct values. If a run container has more than \num{4096}~distinct values, then it must have no more than \num{2047}~runs, otherwise the number of runs must be less than half the number of distinct values.

\begin{figure}
\centering
\begin{tikzpicture}[thick,scale=0.8, every node/.style={transform shape},node distance=0cm,start chain=1 going below,
start chain=2 going below,
start chain=3 going below,
start chain=9 going right,start chain=13 going right,start chain=15 going right,start chain=17 going right]
 \tikzstyle{mytape}=[draw,minimum height=0.7cm,minimum width=3cm]
\tikzstyle{overbrace style}=[decorate,decoration={brace,raise=2mm,amplitude=3pt}]

\tikzset{%
    position label/.style={%
       above = 3pt,
       text height = 1.5ex,
       text depth = 1ex
    },
   brace/.style={%
     decoration={brace},
     decorate
   }
}

\tikzset{%
    position label/.style={%
       above = 3pt,
       text height = 1.5ex,
       text depth = 1ex
    },
   bracemirror/.style={%
     decoration={brace, mirror},
     decorate
   }
}

    \node(ArrayContainerTop) [on chain=9,mytape,fill=red!20,minimum width=1.5cm] {0x0000};
    \node(ArrayContainerTopLink) [on chain=9,mytape,fill=red!20,minimum width=1.5cm] {};
    \node(RunContainerTop) [on chain=9,mytape,fill=green!20,minimum width=1.5cm] {0x0001};
    \node(RunContainerTopLink) [on chain=9,mytape,fill=green!20,minimum width=1.5cm] {};
    \node(BitsetContainerTop) [on chain=9,mytape,fill=blue!20,minimum width=1.5cm] {0x0002};
    \node(BitsetContainerTopLink) [on chain=9,mytape,fill=blue!20,minimum width=1.5cm] {};

\draw [brace] (ArrayContainerTop.north) -- node [position label, pos=0.5] {16-bit values + pointers to container} (BitsetContainerTopLink.north);

\tikzstyle{arraytape}=[draw,minimum height=0.7cm,,minimum width=2cm,fill=red!20]

    \node (ArrayContainer)[on chain=2,arraytape,below=3cm of ArrayContainerTop] {0};
    \node [on chain=2,arraytape] {62};
    \node [on chain=2,arraytape] {124};
    \node [on chain=2,arraytape] {186} ;
    \node [on chain=2,arraytape] {248};
    \node [on chain=2,arraytape] {310};
    \node [on chain=2,arraytape] {$\vdots$};
    \node (ArrayContainerEnd)[on chain=2,arraytape] {61\,938};

\draw [bracemirror] (ArrayContainer.west) -- node [position label, pos=0.5,rotate=90] {1000~sorted 16-bit values (2000\,B)} (ArrayContainerEnd.west);

\tikzstyle{runtape}=[draw,minimum height=0.7cm,minimum width=2cm,fill=green!20]

    \node (RunContainer)[on chain=2,runtape,below=3cm of RunContainerTop]{$\langle 0,99\rangle$}; 
\node [on chain=2, runtape] {$\langle 101,99\rangle$};
\node (RunContainerEnd)[on chain=2, runtape] {$\langle 300,99\rangle$};

\draw [brace] (RunContainerEnd.south east) -- node [position label, pos=0.5, below=40pt] {\parbox{2.5cm}{Three pairs of 16-bit values, indicating the intervals $[0,99],[101,200],$ $[300,399] $; pairs are sorted for fast random access.}} (RunContainerEnd.south west);

\draw [brace] (RunContainer.east) -- node [position label, pos=0.5,rotate=270] {$3\times$ 2~B$= 6$~B} (RunContainerEnd.east);

\tikzstyle{bitsettape}=[draw,minimum height=0.7cm,,minimum width=2cm,fill=blue!20]

   \node (BitsetContainer)[on chain=3,bitsettape,below=3cm of BitsetContainerTop] {1};
    \node [on chain=3,bitsettape] {0};
    \node [on chain=3,bitsettape] {1};
    \node [on chain=3,bitsettape] {0} ;
    \node [on chain=3,bitsettape] {1};
    \node [on chain=3,bitsettape] {0};
    \node [on chain=3,bitsettape] {$\vdots$};
    \node (BitsetContainerEnd) [on chain=3,bitsettape] {0};

\draw [brace] (BitsetContainer.east) -- node [position label, pos=0.5,rotate=270] {$2^{16}$~bits (8\,kB)} (BitsetContainerEnd.east);

\draw[o->, >=latex', shorten >=2pt, shorten <=2pt, thick, bend right=45,dashed]
    (ArrayContainerTopLink.center) to node[fill=yellow!30] {\parbox{2cm}{array\\ container (1000~values)}}
    (ArrayContainer.north);

\draw[o->, >=latex', shorten >=2pt, shorten <=2pt, thick, bend right=45,dashed]
    (BitsetContainerTopLink.center) to node[fill=yellow!30] {\parbox{2cm}{bitset\\ container  ($2^{15}$~values)}}
    (BitsetContainer.north);

\draw[o->, >=latex', shorten >=2pt, shorten <=2pt, thick, bend right=45,dashed]
    (RunContainerTopLink.center) to node[fill=yellow!30] {\parbox{2cm}{run\\ container  (3~runs)}}
    (RunContainer.north);

\end{tikzpicture}
\caption{\label{fig:roaringbitmap} Roaring bitmap containing the first 1000~multiples of 62,   all integers in the intervals
$[2^{16},2^{16}+100)$,
$[2^{16}+101,2^{16}+201)$,
$[2^{16}+300,2^{16}+400)$  and  all even integers in $[2\times 2^{16},3\times 2^{16})$.}
\end{figure}

Roaring offers logarithmic-time random access: to check for the presence of a 32-bit integer, we seek the container corresponding to the sixteen most-significant bits using a binary search.
If this prefix is not in the list, we know that the integer is not
present.  If a bitmap container is found, we check the corresponding bit; if an array or run container is found, we use a binary search.

The early implementations of Roaring are in Java~\cite{SPE:SPE2325,RoaringDocIdSet}. Though Java offers high performance, it also abstracts away the hardware, making it more difficult for the programmer to tune the software to the microarchitecture.
As an experiment, we built an optimized Roaring bitmap library in C (CRoaring). Based on this work, we make two main contributions:
\begin{itemize}
\item We present several non-trivial algorithmic optimizations. See Table~\ref{tab:optimizations}. In particular, we show that
a collection of algorithms exploiting
 single-instruction-multiple-data (SIMD) instructions can enhance the performance of a data structure like Roaring in some cases, above and beyond what state-of-the-art optimizing compilers can achieve. To our knowledge, it is the first work to report on the benefits of advanced SIMD-based algorithms for compressed bitmaps.

 Though the approach we use to compute array intersections using SIMD instructions in \S~\ref{sec:simdinter} is not new~\cite{simdcompandinter,Schlegel2011}, our work on  the computation of the union (\S~\ref{sec:simdunion}), difference (\S~\ref{sec:simddifference}) and symmetric difference (\S~\ref{sec:simddifference}) of arrays using SIMD instructions might be novel and of general interest. 
\item We benchmark our C library against a wide range of alternatives in C and C++. Our results  provide guidance as to the strengths and weaknesses of our implementation.
\end{itemize}
We focus primarily on our novel implementation and the lessons we learned:
we refer to earlier work for details regarding the high-level algorithmic design of Roaring bitmaps~\cite{SPE:SPE2325,lemire2016consistently}. Because our library is freely available under a liberal open-source license, we hope that our work will be used to accelerate information systems.

\begin{table}[h]
\caption{\label{tab:optimizations}Our Roaring Optimizations and how they relate to prior work. }
\centering
\begin{tabular}{p{0.2\textwidth}p{0.4\textwidth}cc}
\toprule
containers &optimization  & section & prior work\\
        \midrule
bitset$\to$array & converting bitsets to arrays & \S~\ref{sec:convertbitset} & \\
bitset+array & setting, flipping or resetting the bits of a bitset at indexes specified by an array, with and without cardinality tracking & \S~\ref{sec:arraybitsetaggregates}& \\
bitset   &  computing the cardinality using a vectorized Harley-Seal algorithm & \S~\ref{sec:harleyseal} & \cite{mulapopcount} \\
bitset+bitset   &  computing AND/OR/XOR/ANDNOT between two bitsets with cardinality using a vectorized Harley-Seal algorithm & \S~\ref{sec:vecopoverbitsets}&  \cite{mulapopcount} \\
array+array  & computing the intersection between two arrays using a vectorized algorithm & \S~\ref{sec:simdinter} & \cite{simdcompandinter,Schlegel2011} \\
array+array  & computing the union between two arrays using a vectorized algorithm & \S~\ref{sec:simdunion}&  \\
array+array  & computing the difference between two arrays using a vectorized algorithm & \S~\ref{sec:simddifference} & \\
array+array  & computing the symmetric difference between two arrays using a vectorized algorithm & \S~\ref{sec:simdsymmetricdifference}&  \\
\bottomrule
\end{tabular}
\end{table}

\section{Integer-Set Data Structures}
\label{sec:rel}

The simplest way to represent a set of integers is as a sorted array
, leading to an easy implementation.
 Querying for the presence of a given value can be done in logarithmic time using a binary search. Efficient set operations are already supported in standard libraries (e.g., in C++ through the Standard Template Library  or STL). We can compute the intersection, union, difference, and symmetric difference between two sorted arrays in linear time: $O(n_1 + n_2)$ where $n_1$ and $n_2$ are the cardinalities of the two arrays. The intersection and difference can also be computed in time $O(n_1 \log n_2)$, which is advantageous when one array is small.

Another conventional approach is to implement a set of integers using a hash set---effectively a hash table without any values attached to the keys. Hash sets are supported in the standard libraries of several popular languages  (e.g., in C++ through  \texttt{unordered_set}). Checking for the presence of a value in a hash set can be done in expected constant time,
giving other hash-set operations favorable computational complexities. 
For example, adding a set of size $n_2$ to an existing set of size $n_1$ can be done in expected linear time $O(n_2)$---an optimal complexity. The intersection between two sets can be computed in expected $O(\min (n_1, n_2))$ time.  However, compared to a sorted array, a hash set is likely to use more memory. Moreover, and maybe more critically, accessing data in the hash set involves repeated  random
accesses to memory, instead of the more efficient sequential access made
possible by a sorted array.

\subsection{Compressed Bitsets}
\label{sec:compressedbitsets}
A bitset (or bitmap) has both the benefits of the hash set (constant-time random access) and of a sorted array (good locality), but suffers from impractical 
 memory usage when the universe size is too large compared to the cardinality of the sets. So we use compressed bitmaps.
Though there are alternatives~\cite{navarro2012fast}, the most
popular bitmap compression techniques are based on the word-aligned RLE
compression model inherited from Oracle (BBC~\cite{874730}):
WAH~\cite{wu2008breaking},
Concise~\cite{colantonio:2010:ccn:1824821.1824857},
EWAH~\cite{arxiv:0901.3751} (used by the Git~\cite{gitewah} version control system), COMPAX~\cite{netfli},
VLC~\cite{Corrales:2011:VLC:2033546.2033586},
VAL-WAH~\cite{guzuntunable}, among others~\cite{7249457,7448840,7442497}.
On a $W$-bit system, the $r$~bits of the bitset are partitioned into sequences of $W'$ consecutive bits,
where $W' \approx W$ depends on the technique used; for
EWAH, $W' = W$; for WAH and many others, $W'=W-1$.
When such a sequence of $W'$ consecutive bits contains only 1s or only 0s,
we call it a \emph{fill}~word,
otherwise
we call it a  \emph{dirty} word.
For example, using $W'=8$, the uncompressed bitmap $00000000 01010000$ contains two words,
a fill word ($00000000$) and a dirty word ($01010000$).
Techniques such as BBC, WAH or EWAH  use special
marker words to compress long sequences of identical
{fill}~words.
When accessing these formats, it may be necessary to read
every compressed word to determine whether it indicates a sequence of fill words,
or a
{dirty}~word.
The EWAH format supports a limited form of skipping because
its marker words not only to give the length of the sequences of
fill words, but  also  the number of consecutive dirty words.
EWAH was found to have superior performance to WAH and Concise~\cite{guzuntunable} in an extensive comparison.
A major limitation of formats like BBC, WAH, Concise or EWAH is that random access is slow. That is, to check whether a value is present in a set can take linear time $O(n)$, where $n$ is the compressed size of the bitmap. When intersecting a small set with a more voluminous one, these formats may have  suboptimal performance.

Guzun et al.~\cite{Guzun2016} compared Concise, WAH, EWAH, Roaring (without run containers) with a hybrid system combining EWAH and uncompressed bitsets. In many of their tests (but not all), Roaring was superior to  Concise, WAH, and EWAH\@.
Similarly, Wang et al.~\cite{goroaring2017} compared a wide range of bitmap compression techniques. They conclude their comparison by a strong recommendation in favor of Roaring: ``Use Roaring for bitmap compression whenever possible. Do not use other bitmap compression methods such as BBC, WAH, EWAH, PLWAH, CONCISE, VALWAH, and SBH.''

\subsection{BitMagic}
\label{sec:bitmagic}

Mixing container types in one data structure is not unique to Roaring. O'Neil and O'Neil's RIDBit is an external-memory B-tree of  bitsets and lists~\cite{O'Neil:2007:BID:1304611.1306564,Rinfret:2001:BIA:375663.375669}.
Culpepper and Moffat's \textsc{hyb+m2} mixes compressed arrays with bitmaps~\cite{culpepper:2010:esi:1877766.1877767}---Lemire et al.~\cite{simdcompandinter} decomposed \textsc{hyb+m2} into chunks of bitsets and arrays that fit in CPU cache.

The BitMagic library is probably the most closely related data structure~\cite{BitMagiclib} with a publicly available implementation. Like Roaring, it is a two-level data structure similar to RIDBit~\cite{O'Neil:2007:BID:1304611.1306564,Rinfret:2001:BIA:375663.375669}.
There are, however, a few differences between Roaring and BitMagic:
\begin{itemize}
\item
In BitMagic, special pointer values are used to indicate a full (containing $2^{16}$~values) or an empty container; Roaring does not need to mark empty containers (they are omitted) and it can use a run container to represent a full container efficiently.
\item Roaring relies on effective heuristics to generate a memory-efficient container. For example, when computing the union between two array containers, we 
guess whether the output is likely to be more efficiently represented as a bitset container, as opposed to an array. See Lemire et al.~\cite{lemire2016consistently} for a detailed discussion. BitMagic does not attempt to optimize the resulting containers: e.g., the intersection of two bitset containers is a bitset container, even when another container type could be more economical.

Roaring keeps 
a cardinality counter updated
for all its bitset containers. BitMagic keeps track of the set cardinality, but not at the level of the bitset containers. Hence, whereas deleting a value in a bitset container in Roaring might force a conversion to an array container, BitMagic cannot know that a conversion is necessary without help: the user must manually call the  \texttt{optimize} method.

When aggregating bitmaps, it is not always useful to maintain the cardinality of the intermediate result: for this purpose, Roaring allows specific \emph{lazy} operations. The BitMagic library, instead, uses a flag to indicate whether the  precomputed cardinality is valid, and systematically invalidates it following any aggregation operation.
Thus, by default, BitMagic operations can be faster because the cardinality of the bitset containers is not tracked, but the resulting bitmaps may not be compressed as well as they could be: bitset containers might be used to store few values.

Intuitively, one might think that tracking the cardinality of a bitset as we operate on it---as we do in Roaring by default---could be  expensive. However, as we show in \S~\ref{sec:bitmanip} and \S~\ref{sec:vecproc}, careful optimization can make the overhead of tracking the cardinality small on current processors.

\item
Roaring uses a key-container array, with one entry per  non-empty container; BitMagic's top-level array has $\lceil n/2^{16}\rceil $~entries to represent a set of values in $[0,n)$. Each entry is made of a pointer to a container. Such a flat layout might  give BitMagic 
a speed advantage over Roaring in some cases.
For example, when doing random-access queries, Roaring relies on a moderately expensive binary search at the top level whereas BitMagic has direct access. Nevertheless, BitMagic's simpler layout has a higher memory usage when many of its containers are  empty.
\item In addition to full and empty containers,
 BitMagic supports two kinds of containers: bitsets and 
``gap'' containers. They are equivalent to Roaring's bitset and run containers, but BitMagic has no equivalent to Roaring's array containers.
Moreover, while both CRoaring and BitMagic use dynamic arrays as containers,  CRoaring attempts to use as little memory as possible, while BitMagic always uses one of a handful of possible run-container sizes (i.e., in 2-byte words: 128, 256, 512 or 1280).
This difference suggests that Roaring may offer better compression
ratios for some datasets. However, it may be easier to provide a custom memory allocator for BitMagic.
\item When appropriate, Roaring uses an intersection approach based on binary search  between array containers instead of systematically relying on a  linear-time merge like RIDBit~\cite{O'Neil:2007:BID:1304611.1306564,Rinfret:2001:BIA:375663.375669}. That is, we use galloping intersection (also known as exponential intersection) to aggregate two sorted arrays of sizes $n_1,n_2$ in linearithmic time ($O(\min(n_1,n_2) \log \max(n_1,n_2))$)~\cite{bentley1976almost}. BitMagic does not use array containers.

 \item
 Unlike the current Roaring implementations, the BitMagic library uses \emph{tagged pointers} to 
distinguish
the container types:  it uses the least significant bit of the address value as a container-type marker bit. In our C implementation, Roaring uses an array of bytes, using one byte per container, to indicate the container type.
\item BitMagic includes algorithms for fast random sampling unlike Roaring.
\end{itemize}
Overall, BitMagic is simpler than Roaring, but we expect that it can sometimes use more memory.

BitMagic includes the following SIMD-based optimizations on x64 processors with support for the SSE2 and SSE4 instruction sets:
\begin{itemize}
\item 
It uses manually optimized SIMD instructions to compute the AND, OR, XOR and ANDNOT operations between
two bitset containers.
\item 
It uses a mix of SIMD instructions and the dedicated population-count instruction (\texttt{popcnt}) for its optimized functions that compute only the cardinality of the result from  AND, OR, XOR and ANDNOT operations between two bitset containers.
\item 
It uses SIMD instructions to negate a bitset quickly.
\end{itemize}

 Naturally, many of our Roaring optimizations (see \S~\ref{sec:vecproc}) would be applicable to BitMagic and similar formats.

\section{Faster Array-Bitset Operations With Bit-Manipulation Instructions}

\label{sec:bitmanip}

Like most commodity processors, Intel and AMD processors benefit from \emph{bit-manipulation instructions}~\cite{wikibmi}. Optimizing compilers often use them, but not always in an optimal manner.

\subsection{Converting Bitsets To Arrays}

\label{sec:convertbitset}

Two useful bit-manipulation instructions are \texttt{blsi}, which  sets all but the least significant 1-bit to zero (i.e., \texttt{x \& -x} in C), and  \texttt{tzcnt}, which counts the number of trailing zeroes (largest $i$ such as that $x/2^i$ is an integer).
Using the corresponding Intel intrinsics (\texttt{_blsi_u64} and \texttt{_mm_tzcnti_64}), we can extract the locations of all 1-bits in a 64-bit word (\texttt{w}) to an array (\texttt{out}) efficiently.
\begin{lstlisting}
pos = 0
while (w != 0) {
  uint64_t temp = _blsi_u64(w);
  out[pos++] = _mm_tzcnti_64(w);
  w ^= temp;
}
\end{lstlisting}
Such code is useful when we need to convert a bitset container to an array container. We can ensure that only a handful of instructions are needed per bit set in the bitset container.

\subsection{Array-Bitset Aggregates}
\label{sec:arraybitsetaggregates}

Several other such instructions are useful when implementing Roaring. For example,  the \texttt{bt} (bittest) instruction returns the value of a bit whereas other bit-manipulation instructions can set (\texttt{bts}), clear (\texttt{btr}) and flip bits (\texttt{btc}) as they query their value. On a Pentium~4, these instructions had a throughput of one instruction every two cycles and a latency of six cycles. However, starting with the Intel Sandy Bridge
microarchitecture (2011), they became much faster
with a throughput of two instructions every cycle and a latency of one cycle~\cite{fog2016instruction} for data in registers. 

Roaring  has bitset containers, which
are implemented as arrays of 64-bit words. Some of the most common operations on bitsets involve getting, setting, flipping or clearing the value of a single bit in one of the words. Sometimes it is also necessary to determine whether the value of the bit was changed.

We are interested in  the following
scenario: given a bitset and an array of 16-bit values, we wish to set (or clear or flip) all bits at the  indexes corresponding to the 16-bit values. For example, if the array is made of the values $\{1,3,7,96,130\}$, then we might want to set the bits at indexes $1,3,7,96,130$ to 1. This operation is equivalent to computing the union between a bitset and an array container.


\begin{itemize}
\item If we do not care about the cardinality, and merely want to set the bit, we can 
use
the simple C expression \texttt{(w[pos >> 6] |= UINT64_C(1) << (pos \& 63)}. Compilers can translate this expression into two shifts (one to identify the word index and one to shift 
the single set bit)
and a bitwise OR.

On a recent x64 processor (see \S~\ref{sec:hardware}) and using a recent popular C compiler (see \S~\ref{sec:software}), we estimate that we can set one bit every $\approx 3.6$~cycles, taking into account the time required to read the positions of the bits to be set from an array.

We can do better by using a single shift (to identify the word index) followed by
the instruction \texttt{bts} (``bit test and set'').
Indeed, if the position of the bit in the bitset is given by the \texttt{pos} variable, then we shift \texttt{pos} right by 6 bits and store the result in \texttt{offset}.
With the help of \texttt{offset}, we fetch the affected word from memory
into a variable \texttt{load}.
We call \texttt{bts} with \texttt{pos} and \texttt{load}. This sets the corresponding bit in \texttt{load}, so we can store it back in memory at index \texttt{offset}. Thus we generate a total of four instructions (one load, one shift, one \texttt{bts} and one store)---assuming that the position of the bit to set is already in a register. We find that we are able to set bits about twice as fast (one every $\approx 1.7$~cycles).
 We benefit from the fact that processors are superscalar, so that they can execute more than one instruction per cycle.

%
\item

We could set bits and  track the cardinality
but
we expect 
branch predictions to be sometimes difficult, and branch mispredictions can be expensive (e.g., over 10~cycles of penalty per misprediction).
 Using the following C code, we set bits and track the cardinality while avoiding branches.
\begin{lstlisting}
uint64_t old_w = words[pos >> 6];
uint64_t new_w = old_w | (UINT64_C(1) << (pos & 63));
cardinality += (old_w ^ new_w) >> (pos & 63);
words[pos >> 6] = new_w;
\end{lstlisting}
This code first retrieves the word \texttt{old\_w} at index \texttt{pos/64} (or equivalently \texttt{pos >> 6}) in an array of 64-bit words \texttt{words}.
We then set the bit at index \texttt{pos \% 64} (or equivalently \texttt{pos \& 63}) to 1, and call the result \texttt{new\_w}.
We can write back \texttt{new\_w} to the array \texttt{words}, and use the bitwise XOR 
between 
\texttt{old\_w} 
and 
(\texttt{new\_w}) to determine whether the bit value was changed.
We get about $\approx 3.6$~cycles per bit set, or about the same speed we get when we use similar C code without tracking the cardinality.

We can also achieve the same result, set the bit, and adjust the cardinality counter 
as follows:  Use
a shift (to identify the word index),
the instruction \texttt{bts}, 
then
an instruction like \texttt{sbb} (``integer subtraction with borrow'') to modify the cardinality counter according to the carry flag eventually set by \texttt{bts}. Indeed, if the position of the bit in the bitset is given by the \texttt{pos} variable, then we can proceed as follows:
\begin{itemize}
\item We shift \texttt{pos} right by 6 bits (e.g., \texttt{shrx}) and store the result in \texttt{offset}.
\item We load the 64-bit word at index \texttt{offset} and store it in \texttt{load}.
\item We call \texttt{bts} with \texttt{pos} and \texttt{load}. This sets the corresponding bit in \texttt{load}, so we can store it back in memory at index \texttt{offset}.
\item The call to \texttt{bts} stored the value of the bit  in the  carry (CF) flag. If the bit was set, then the carry flag has value 1, otherwise it has value 0. We want to increment our counter when the carry flag has value 1, and leave it unchanged when it has value 0. Thus we can call the subtract with borrow (\texttt{sbb}) instruction, passing our cardinality counter as well as the parameter -1. It will add -1 to the carry flag (getting 0 and -1), and then subtract this value from the counter, effectively adding 0 or 1 to it.
\end{itemize}
In total, we need five instructions (one load, one shift, one \texttt{bts}, one store and one \texttt{sbb}), which is one more instruction than if we did not have to track the cardinality.

With this approach, we find that we can set a bit while maintaining a cardinality counter every $\approx 1.8$~cycles on our processor. It is only 0.1~cycles per bit set slower than if we do not track the cardinality.

\end{itemize}


According to our experiments, you can set a bit, or set a bit while maintaining the cardinality, with almost  the same  throughput---even when all the data is in 
L1 processor
cache.
 A similar conclusion holds for bit flipping or bit clearing.
Thus it is inexpensive to track the cardinality of bitset containers as we modify individual bits on recent x64 processors.

We also find that if we use bit-manipulation instructions and hand-tuned assembly, we can roughly double the speed at which we can change bits in a bitset, compared to compiler-generated machine code with all optimization flags activated.
It is true not only when setting bits, but also when clearing or flipping them---in which cases we need to use the \texttt{btr} and \texttt{btc} instructions instead. Informally, we tested various C compilers and could not find any that could perform nearly as well as hand-tuned assembly code.
This may, of course, change in the future.

\section{Vectorized Processing}
\label{sec:vecproc}

Modern commodity processors use parallelism to accelerate processing.
SIMD instructions offer a particular form of processor
parallelism~\cite{5009071} that proves advantageous for processing
large volumes of data~\cite{sparkSIMD}. Whereas regular instructions operate on a
single machine word (e.g., 64~bits), SIMD instructions operate on
large registers (e.g., 256~bits) that can be used to represent
``vectors'' containing several distinct values. For example,
a single SIMD instruction can add sixteen 16-bit integers in one
\SI{32}{B} vector register to the corresponding 16-bit integers in another
vector register using a single cycle.  Moreover, modern processors are
capable of superscalar execution of certain vector instructions, allowing the execution of several SIMD instructions per cycle.
For example, an Intel Skylake processor is capable of
executing two vector additions every cycle~\cite{fog2016instruction}.

SIMD instructions are ideally suited for operations between bitset containers. When computing the intersection, union, difference or symmetric difference between two (uncompressed) bitsets, we only have to load the data in registers, apply a logical operation between two words (AND, OR, AND NOT, or XOR) and, optionally, save the result to memory. All of these operations have corresponding SIMD instructions. So, instead of working over 64-bit words, we work over larger words (e.g., 256~bits), dividing the number of instruction (e.g., $\div 4$) and giving 
significantly faster processing.

Historically, SIMD instructions have gotten wider and increasingly powerful. The Pentium~4 was limited to 128-bit instructions. Later x64 processors introduced increasingly powerful 128-bit instruction sets: SSSE3, SSE4, SSE4.1, SSE4.2.
Today, we find rich support for 256-bit vector registers in the AVX2
instruction set available in recent x64 processors from Intel
(starting with the Haswell microarchitecture, 2013)
and AMD (starting
with the Excavator microarchitecture, 2015).
Upcoming Intel commodity processors will support 512-bit SIMD instructions (AVX-512). In many cases, the benefits of SIMD instructions increase as they get wider, which means that SIMD instructions tend to become more compelling with time.

Compilers can
automatically translate C or C++ code into the appropriate SIMD
instructions, even when the code does not take into
account vectorization (scalar code). However, for greater gains, it is
also possible to design algorithms and code that take into account the
vector instructions. In these instances, we can use the SIMD
intrinsics available in C and C++ to call SIMD instructions without
having to use assembly code. See Table~\ref{ref:avxinstructions}. These intrinsics are supported across several compilers (LLVM's Clang, GNU GCC, Intel's compiler, Visual Studio).

\begin{table}[bt]
\caption{Some relevant AVX and AVX2 instructions  with reciprocal throughput$^{\mathrm{a}}$\tnote{a} in CPU cycles on recent (Skylake) Intel processors\label{ref:avxinstructions}.}\centering\scriptsize
\begin{threeparttable}[b]
\begin{tabular}{ccp{1.7in}c} 
\toprule
instruction & C intrinsic & description&  \multicolumn{1}{p{0.6in}}{rec.\ throughput~\cite{fog2016instruction}}
\\\midrule
\texttt{vpand} & \texttt{_mm256_and_si256} & 256-bit AND  & 0.33\\
\texttt{vpor} & \texttt{_mm256_or_si256} & 256-bit OR & 0.33\\
\texttt{vpxor} & \texttt{_mm256_xor_si256} & 256-bit XOR  & 0.33\\
\texttt{vpsadbw} & \texttt{_mm256_sad_epu8} & sum of the absolute differences of the byte values to the low 16 bits of each 64-bit word & 1\\
 \texttt{vpaddq} & \texttt{_mm256_add_epi64} & add 64-bit integers & 0.5\\
 \texttt{vpshufb} & \texttt{_mm256_shuffle_epi8} & \emph{shuffle}  bytes within 128-bit lanes  &  1\\
\bottomrule
\end{tabular}
\begin{tablenotes}
\item [a] The reciprocal throughput is the number of processor clocks it takes for an instruction to execute. A reciprocal throughput of 0.5 means that we execute two such instructions per clock.
\end{tablenotes}
\end{threeparttable}

\end{table}

\subsection{Vectorized Population Count Over Bitsets}
\label{sec:vecpopcnt}

We can trivially vectorize operations between bitsets. Indeed, it
suffices to compute bitwise operations over vectors instead of machine
words. By aggressively unrolling the resulting loop, we can produce
highly efficient code. Optimizing compilers can often automatically vectorize such code.
 It is more difficult, however, to also
compute the cardinality of the result efficiently.  Ideally, we would like to vectorize simultaneously the operations between bitsets and the computation of the cardinality of the result.

\subsubsection{Vectorized Harley-Seal Population Count}
\label{sec:harleyseal}
Commodity processors have
dedicated instructions to count the  1-bits in a word (the
``population count''): \texttt{popcnt} for x64 processors and
\texttt{cnt} for the 64-bit ARM architecture. On recent Intel
processors, \texttt{popcnt} has a throughput of one instruction per
cycle for both 32-bit and 64-bit registers~\cite{fog2016instruction}.

 Maybe surprisingly, Muła et al.\ found that it is possible to do
better~\cite{mulapopcount}. We can use a vectorized approach based on
the circuit-based Harley-Seal algorithm~\cite{warren2007}. It is
inspired by a carry-save adder (CSA) circuit: given 3~bit values
($a,b,c\in \{0,1\}$), we can sum them up into the two-bit value $(a \text{  XOR } b)
\text{ XOR } c$ (least significant) and $(a \text{ AND } b)
\allowbreak \text{ OR } \allowbreak ( (a \text{ XOR } b) \text{ AND } c )$ (most
significant). We can sum 3~individual bit values to a 2-bit counter using
5~logical operations, and we
can generalize this approach to 256-bit
vectors. Starting with 3~input vectors, we can generate two new
output vectors, one containing the least significant bits and one
containing the most significant (or \emph{carry}) bits with 5~bitwise
logical operations (two XORs, two ANDs and one OR):

\begin{lstlisting}
void CSA(__m256i *h, __m256i *l, __m256i a,
  __m256i b, __m256i c) {
  // starts with a,b,c, sets H,L, u is temp.
  __m256i u = _mm256_xor_si256(a, b);
  *h = _mm256_or_si256(_mm256_and_si256(a, b),
    _mm256_and_si256(u, c));
  *l = _mm256_xor_si256(u, c);
}
\end{lstlisting}

From such an adder
function, we can derive an efficient population-count function,
effectively composing 3-input adders into a 
circuit with 16 inputs, which encodes the population count
of its 16 inputs as a 5-bit binary number.
For instance,  Knuth presents a
construction~\cite{KnuthV4A} requiring only 63 logical operations
in the case of a 16-input circuit.
Although using slightly more logical operations (75),
the iterated Harley-Seal approach goes further: it increments
the existing contents of a 5-bit binary number by the population count.

In a SIMD setting, our Harley-Seal circuit takes sixteen 
256-bit vectors
and updates five 
256-bit vectors serving as a bit-sliced
accumulator~\cite{253268}: in other words, if one counts the number
of 1s amongst the least-significant bits of the 16 inputs, the
population count (0 to 16) affects the least-significant bits of the 5
accumulator vectors.  Similarly, the population count of the
second-least-significant bits of the 16 inputs affects the
second-least-significant bits of the accumulator vectors.
Equivalently, the 5 accumulator vectors can be viewed as
providing 256 different 5-bit accumulators.  The
accumulator vectors are named \texttt{ones}, \texttt{twos},
\texttt{fours}, \texttt{eights}, and \texttt{sixteens},
reflecting their use in the accumulator: if the least significant bits of
the accumulator vectors are all initially zero and the population count of the
least-significant bits of the 16 input vectors is 9, then our
Harley-Seal approach will result in the least-significant bits of
\texttt{eights} and \texttt{ones} being set, whereas the
least-significant bits of \texttt{sixteens}, \texttt{fours} and
\texttt{twos} would be clear.  (The accumulator went from storing 0 to
storing 9.)  If the accumulator updated a second time with the same
inputs, then the least-significant bits of \texttt{sixteens} and
\texttt{twos} (only) would be set.


Fig.~\ref{fig:hsillustration} (from~\cite{mulapopcount}) illustrates a simplified case with only \texttt{ones}, \texttt{twos} and \texttt{fours}.
Fig.~\ref{fig:vectorized-popcount} shows the high-level process:
We load 16~vectors from the 
source bitset
, we compute the bitwise operations,
and 
obtain the result in
 \texttt{ones}, \texttt{twos}, \texttt{fours}, \texttt{eights} and \texttt{sixteens}.
%
This processes a \SI{4}{kb} block of an input bitset using
approximately 75~AVX operations, not considering any data-movement
instructions (see Fig.~\ref{code:circuit}). 
We can compute the total value of all 256~accumulators as
$16\times\mathrm{count}(\texttt{sixteens})+8\times\mathrm{count}(\texttt{eights})+4\times\mathrm{count}(\texttt{fours})+2\times \mathrm{count}(\texttt{twos})+\mathrm{count}(\texttt{ones})$ (13~more operations),
where $\mathrm{count}(b)$ denotes
the population count of $b$.
Iterated 16 times, we can process the entire input bitset.  Moreover, it is only necessary to read the total value
of all 256 accumulators after the last iteration.
However, we need to ensure that, at the beginning of each iteration, none of our 5-bit accumulators ever begins with a count
above 15 --- otherwise, an accumulator might need to store a value exceeding 31.
Therefore,
at the end of each block we count the 1-bits in each of the four~64-bit words  of \texttt{sixteens} and add the result to a vector counter $c$ before zeroing \texttt{sixteens}.
To count the 1-bits in \texttt{sixteens} quickly, we use vector registers as lookup tables mapping 4-bit values (integers in $[0,16)$) to their corresponding population counts, and the \texttt{vpshufb} can effectively look up 32~byte values at once. Thus we can divide each byte of \texttt{sixteens} into its low nibble and high nibble and look up the population count of each. To get a population count for the 64-bit words, we use the instruction \texttt{vpsadbw} (with the intrinsic \texttt{_mm256_sad_epu8}): this instruction adds the absolute values of the differences between byte values within 64-bit subwords. We illustrate the code in Fig.~\ref{fig:avxmula}: the vectors \texttt{lookuppos} and \texttt{lookupneg} store the positives and negatives of the population counts with a canceling offset of 4 so that their subtraction is the sum of the population counts.
After the $16^{\textrm{th}}$ iteration, the population count of the entire input bitset is computed as
$16\times c+8\times\mathrm{count}(\texttt{eights})+4\times\mathrm{count}(\texttt{fours})+2\times \mathrm{count}(\texttt{twos})+\mathrm{count}(\texttt{ones})$.

\begin{figure}\centering
\begin{tabular}{c}
\begin{lstlisting}
__m256i popcount256(__m256i v) {
  __m256i lookuppos = _mm256_setr_epi8(
        4 + 0, 4 + 1, 4 + 1, 4 + 2,
        4 + 1, 4 + 2, 4 + 2, 4 + 3,
        4 + 1, 4 + 2, 4 + 2, 4 + 3,
        4 + 2, 4 + 3, 4 + 3, 4 + 4,
        4 + 0, 4 + 1, 4 + 1, 4 + 2,
        4 + 1, 4 + 2, 4 + 2, 4 + 3,
        4 + 1, 4 + 2, 4 + 2, 4 + 3,
        4 + 2, 4 + 3, 4 + 3, 4 + 4);
  __m256i lookupneg = _mm256_setr_epi8(
        4 - 0, 4 - 1, 4 - 1, 4 - 2,
        4 - 1, 4 - 2, 4 - 2, 4 - 3,
        4 - 1, 4 - 2, 4 - 2, 4 - 3,
        4 - 2, 4 - 3, 4 - 3, 4 - 4,
        4 - 0, 4 - 1, 4 - 1, 4 - 2,
        4 - 1, 4 - 2, 4 - 2, 4 - 3,
        4 - 1, 4 - 2, 4 - 2, 4 - 3,
        4 - 2, 4 - 3, 4 - 3, 4 - 4);
  __m256i low_mask = _mm256_set1_epi8(0x0f);
  __m256i lo = _mm256_and_si256(v, low_mask);
  __m256i hi = _mm256_and_si256(_mm256_srli_epi16(v, 4),
    low_mask);
  __m256i popcnt1 = _mm256_shuffle_epi8(lookuppos, lo);
  __m256i popcnt2 = _mm256_shuffle_epi8(lookupneg, hi);
  return _mm256_sad_epu8(popcnt1, popcnt2);
}
\end{lstlisting}
\end{tabular}
\caption{\label{fig:avxmula}A C function using AVX2 intrinsics to compute the four population counts of the four 64-bit words in a 256-bit vector.}
\end{figure}


By contrast, in the approach using the dedicated population-count instruction, the compiler generates one load, one \texttt{popcnt} and one \texttt{add} per input 64-bit word.
Current x64 processors
decompose complex machine instructions into low-level instructions called \microops{}.
The \texttt{popcnt} approach generates three \microops{} per word. For the SIMD approach, we process sixteen 256-bit vectors using 98~\microops{} including 16~loads, 32~bitwise ANDs (\texttt{vpand}), 15~bitwise ORs (\texttt{vpor}) and 30~bitwise XORs (\texttt{vpxor})---or about 1.5~\microops{} 
to process each
64-bit word. That is, the vectorized approach generates half the number of~\microops{}. We analyzed our implementations with the IACA code analyser~\cite{intelIACA} for our target x64 microarchitecture (\S~\ref{sec:hardware}).  Unsurprisingly, IACA predicts a throughput of one word per cycle with the dedicated population-count function, and a throughput of slightly more than two words per cycle with the vectorized approach.
Microbenchmarking agrees with IACA and shows that our approach is roughly
twice as fast as relying exclusively on dedicated population-count instructions.

\begin{figure}
\begin{center}
\begin{tabular}{c}
\begin{lstlisting}
CSA(&twosA,&ones,ones,A[i],A[i + 1]);
CSA(&twosB,&ones,ones,A[i + 2],A[i + 3]);
CSA(&foursA,&twos,twos,twosA,twosB);
CSA(&twosA,&ones,ones,A[i + 4],A[i + 5]);
CSA(&twosB,&ones,ones,A[i + 6],A[i + 7]);
CSA(&foursB,&twos,twos,twosA,twosB);
CSA(&eightsA,&fours,fours,foursA,foursB);
CSA(&twosA,&ones,ones,A[i + 8],A[i + 9]);
CSA(&twosB,&ones,ones,A[i + 10],A[i + 11]);
CSA(&foursA,&twos,twos,twosA,twosB);
CSA(&twosA,&ones,ones,A[i + 12],A[i + 13]);
CSA(&twosB,&ones,ones,A[i + 14],A[i + 15]);
CSA(&foursB,&twos,twos,twosA,twosB);
CSA(&eightsB,&fours,fours,foursA,foursB);
CSA(&sixteens,&eights,eights,eightsA,eightsB);
\end{lstlisting}
\end{tabular}
\end{center}
\caption{\label{code:circuit}Accumulator circuit over sixteen inputs \texttt{(A[i+0],} \ldots, \texttt{A[i+15])}}
\end{figure}

\begin{figure}
\centering
\begin{tikzpicture}[thick,scale=0.9, every node/.style={transform shape},node distance=0cm,start chain=1 going right,
start chain=9 going right,start chain=10 going right,start chain=11 going right,start chain=12 going right,start chain=13 going right]
 \tikzstyle{mytape}=[minimum height=0.7cm,minimum width=1cm]

    \node(A3) [on chain=9,mytape] {\ldots};
    \node(A4) [on chain=9,mytape] {\parbox{1cm}{\centering \texttt{twos}\\(input)}};
    \node(A5) [on chain=9,mytape] {\parbox{1cm}{\centering \texttt{ones}\\(input)}};
    \node(A6) [on chain=9,mytape] {\parbox{1cm}{\centering $d_i$\\(input)}};
    \node(A7) [on chain=9,mytape]
    {\parbox{1cm}{\centering $d_{i+1}$\\(input)}};

\node(csa1)[on chain=10,fill=gray!30,minimum width=3cm,below=0.5cm of A6] {CSA};
    \node(A8) [on chain=10,mytape]
    {\parbox{1cm}{\centering $d_{i+2}$\\(input)}};
    \node(A9) [on chain=10,mytape]
    {\parbox{1cm}{\centering $d_{i+3}$\\(input)}};
\draw[->,blue, >=latex,thick] (A5.south)  to    (A5.south |- csa1.north);
\draw[->,blue, >=latex,thick] (A6.south)  to    (A6.south |- csa1.north);
\draw[->,blue, >=latex,thick] (A7.south)  to    (A7.south |- csa1.north);
\node(csa2)[on chain=11,fill=gray!30,minimum width=3cm,below=0.5cm of A8] {CSA};
   \node(A10) [on chain=11,mytape] {\ldots};

\draw[->, blue,>=latex,thick]
    ($(csa1.south east)!0.22!(csa1.south)$) to[out=270,in=90]
    ($(csa2.north west)!0.35!(csa2.north)$);
\draw[->, blue,>=latex,thick] (A8.south)  to    (A8.south |- csa2.north);
\draw[->, blue,>=latex,thick] (A9.south)  to    (A9.south |- csa2.north);

\node(csa3)[on chain=12,below=1cm of A10,minimum width=4cm] {};
\node(csa12)[fill=gray!30,minimum width=3cm,left=0.5cm of csa3] {CSA};

\draw[->,red, >=latex,thick]
    (A4.south) to[out=270,in=90]
    ($(csa12.north west)!0.25!(csa12.north)$);

\node(foursoutput)  [on chain=13,mytape,below=1cm of $(csa12.south west)!0.35!(csa12.south)$,minimum width=2cm]
{\parbox{1cm}{\centering \texttt{fours}\\(output)}};

\draw[->,>=latex,thick]
    ($(csa12.south west)!0.35!(csa12.south)$) to[out=270,in=90]
    (foursoutput.north);

\node(twosoutput)  [minimum width=2cm,below=1cm of $(csa12.south east)!0.35!(csa12.south)$]
{\parbox{1cm}{\centering \texttt{twos}\\(output)}};

\draw[->,red, >=latex,thick]
    ($(csa12.south east)!0.35!(csa12.south)$) to[out=270,in=90]
    (twosoutput.north);

\node(onesoutput)  [mytape,minimum width=2cm,below=1.5cm of $(csa2.south east)!0.35!(csa2.south)$]
{\parbox{1cm}{\centering \texttt{ones}\\(output)}};

\draw[->, blue,>=latex,thick]
    ($(csa2.south east)!0.35!(csa2.south)$) to[out=270,in=90]
    (onesoutput.north);

\draw[->,red, >=latex,thick]
    ($(csa2.south west)!0.35!(csa2.south)$) to[out=270,in=90]
    ($(csa12.north east)!0.35!(csa12.north)$);

\draw[->,red, >=latex,thick]
    ($(csa1.south west)!0.35!(csa1.south)$) to[out=270,in=90]
    (csa12.north);
\end{tikzpicture}
\caption{\label{fig:hsillustration}Illustration of the Harley-Seal algorithm aggregating four new inputs ($d_i, d_{i+1}, d_{i+2}$, $d_{i+3}$) to inputs \texttt{ones} and \texttt{twos}, producing \texttt{ones}, \texttt{twos} and \texttt{fours}~\cite{mulapopcount}. }
\end{figure}







\begin{figure}\centering
\begin{tikzpicture}[thick,scale=0.5, every node/.style={transform shape},node distance=0cm,start chain=1 going right,
start chain=9 going right]
 \tikzstyle{mytape}=[draw,minimum height=1.2cm,minimum width=3cm]
 \tikzstyle{mytapew}=[draw,minimum height=2cm,minimum width=3.5cm, fill=magenta!20]

    \node(A1) [on chain=9, mytapew] {\parbox{3.2cm}{\centering\LARGE bits\\ 0 to 255}};
    \node(A2) [on chain=9, mytapew] {\parbox{3.2cm}{\centering\LARGE bits\\  256 to 511}};
    \node(A3) [on chain=9, mytapew] {\parbox{3.2cm}{\centering\LARGE bits\\  512 to 767}};
    \node(A4) [on chain=9, mytapew] {\parbox{3.2cm}{\centering\LARGE bits\\  768 to 1024}};
    \node(A5) [on chain=9,minimum height=0.7cm, minimum width=3cm] {\centering\LARGE \ldots};
    \node(A6) [on chain=9, mytapew] {\parbox{3.2cm}{\centering\LARGE bits\\  3840 to 4095}};
\tikzstyle{overbrace style}=[decorate,decoration={brace,raise=20mm,amplitude=3pt}]

\tikzset{%
    position label/.style={%
       above = 3pt,
       text height = 1.5ex,
       text depth = 1ex
    },
   brace/.style={%
     decoration={brace},
     decorate
   }
}

\draw [brace] (A1.north) -- node [position label, pos=0.5] {\LARGE sixteen 256-bit inputs} (A6.north);

\node(sss)[fill=gray!10,minimum width=25cm,below=2cm of A3.east] {\parbox{15cm}{\centering\LARGE
Harley-Seal accumulator,
}};

\draw[->] (A1.south)  to    (A1.south |- sss.north);
\draw[->] (A2.south)  to    (A2.south |- sss.north);
\draw[->] (A3.south)  to    (A3.south |- sss.north);
\draw[->] (A4.south)  to    (A4.south |- sss.north);
\draw[->] (A6.south)  to    (A6.south |- sss.north);

    \node(B1) [on chain=1,mytape,below=4cm of A1,fill=cyan!20] {\LARGE ones};
    \node(B2) [on chain=1,mytape,fill=cyan!20] {\LARGE twos};
    \node(B3) [on chain=1,mytape,fill=cyan!20] {\LARGE fours};
    \node(B4) [on chain=1,mytape,fill=cyan!20] {\LARGE eights};
    \node(B5) [on chain=1,mytape,fill=cyan!20] {\LARGE sixteens};

\draw[->] (B1.north |- sss.south)  to    (B1.north);
\draw[->] (B2.north |- sss.south)  to    (B2.north);
\draw[->] (B3.north |- sss.south)  to    (B3.north);
\draw[->] (B4.north |- sss.south)  to    (B4.north);
\draw[->] (B5.north |- sss.south)  to    (B5.north);
\end{tikzpicture}
\caption{\label{fig:vectorized-popcount}Vectorized population count for a \SI{4}{kb} bitset,
folding sixteen 256-bit inputs into five 256-bit outputs.}
\end{figure}

\subsubsection{Vectorized Operations Over Bitsets With Population Counts}
\label{sec:vecopoverbitsets}
 To aggregate two bitsets (e.g., with the intersection/AND, union/OR, etc.) while
maintaining the cardinality, we found that an efficient approach is to
load two machine words from each bitset, compute the bitwise
operation, apply the population-count instruction on the two resulting
words, and write them out before processing the next words. To benefit from vectorization, we 
could 
compute the bitwise logical operations using SIMD instructions, and then reload the result from memory and apply the population-count instructions. The interplay between SIMD instructions and a population-count function operating over 64-bit words 
would not be
ideal.

Though an obvious application of the vectorized population count is the efficient computation of the cardinality of a bitset container, we put also it to good use whenever we need to compute the intersection, union, difference or symmetric difference between two bitset containers, while, at the same time, gathering the cardinality of the result. We can also use it to compute the cardinality of the intersection, union, difference or symmetric difference, without materializing them.  As reported by Muła et al.~\cite{mulapopcount}, in these cases, the benefits of the vectorized approach can exceed a factor of two.
Indeed, if we use a vectorized approach to the population count, we can count the bits directly on the vector registers, without  additional data transfer.

\subsection{Vectorized Intersections  Between Arrays}
\label{sec:simdinter}
Because array containers represent integers values as sorted arrays of
16-bit integers, we can put to good use an algorithm based on a
vectorized string comparison function present in recent x64 processors~\cite{simdcompandinter,Schlegel2011} (SSE~4.1's \texttt{pcmpistrm}).
While intended for string comparisons, we can
repurpose the instruction to compute intersections.
Each
input array is divided into blocks of eight 16-bit integers. There are
two block pointers (one for each array) initialized to the first
blocks. We use the \texttt{pcmpistrm} instruction to conduct an
all-against-all comparison of 16-bit integers between these two blocks
and we extract matching integers from a mask produced by the
instruction.  The number of 1s in the resulting mask indicates the
number of matching integers.
To avoid the cost of comparing pairs of blocks that cannot match,
a merge-like process considers the maximum element
within each block to find those block-pairs to which
\texttt{pcmpistrm} should be applied. That is,
whenever the current maximum value in the block of one array is smaller or equal to the maximum of the other block, we load the next block. In particular, when the maxima of the two blocks are equal, we load a new block in each array. See Algorithm~\ref{algo:geninteralgo}.

\begin{algorithm}
\begin{algorithmic}[1]
\REQUIRE two non-empty arrays made of a multiple of $K$ distinct values in sorted order
\STATE load a block $B_1$ of $K$ elements from the first array
\STATE load a block $B_2$ of $K$ elements from the second array
\WHILE {true}
\STATE write the intersection between the two blocks $B_1\cap B_2$ to the output
\IF{ $\max B_1 = \max B_2$}
\STATE load a new block $B_1$ of $K$ elements from the first array, or terminate if exhausted
\STATE load a new block $B_2$ of $K$ elements from the second array, or terminate if exhausted
\ELSIF{$\max B_1 < \max B_2$}
\STATE load a new block $B_1$ of $K$ elements from the first array, or terminate if exhausted
\ELSIF{$\max B_2 < \max B_1$}
\STATE load a new block $B_2$ of $K$ elements from the second array, or terminate if  exhausted
\ENDIF
\ENDWHILE
\end{algorithmic}
\caption{Generic block-based intersection algorithm\label{algo:geninteralgo}.}
\end{algorithm}

We illustrate a simplified function in Fig.~\ref{code:simpinter}. The \texttt{_mm_cmpistrm} intrinsic compares all 16-bit integers from one 128-bit vector with all 16-bit integers from another vector, returning a mask. From this mask, we compute the number of matching values (using the \texttt{_mm_popcnt_u32} intrinsic to call the \texttt{popcnt} instruction). Moreover, we can shuffle the values in one of the vectors (\texttt{vA}) using a shuffling intrinsic (\texttt{_mm_shuffle_epi8}) and a table of precomputed shuffling masks (\texttt{mask16}).  
%
%
We could design an equivalent AVX2 function~\cite{simdcompandinter} using 256-bit vectors, but AVX2 lacks the equivalent of the  \texttt{_mm_cmpistrm} intrinsic.


\begin{figure}
\begin{lstlisting}
int32_t intersect(uint16_t *A, size_t lengthA,
    uint16_t *B, size_t lengthB,
    uint16_t *out) {
  size_t count = 0; // size of intersection
  size_t i = 0, j = 0;
  int vectorlength = sizeof(__m128i) / sizeof(uint16_t);
  __m128i vA, vB, rv, sm16;
  while (i < lengthA) && (j < lengthB) {
    vA = _mm_loadu_si128((__m128i *)&A[i]);
    vB = _mm_loadu_si128((__m128i *)&B[j]);
    rv = _mm_cmpistrm(vB, vA,
      _SIDD_UWORD_OPS | _SIDD_CMP_EQUAL_ANY | _SIDD_BIT_MASK);
    int r = _mm_extract_epi32(rv, 0);
    sm16 = _mm_load_si128(mask16 + r);
    __m128i p = _mm_shuffle_epi8(vA, sm16);
    _mm_storeu_si128((__m128i *)(out + count),p);
    count += _mm_popcnt_u32(r);
    uint16_t a_max = A[i + vectorlength - 1];
    uint16_t b_max = B[j + vectorlength - 1];
    if (a_max <= b_max) i += vectorlength;
    if (b_max <= a_max) j += vectorlength;
  }
  return count;
}
\end{lstlisting}
\caption{\label{code:simpinter}Optimized intersection function between two sorted 16-bit arrays using SSE4.1 intrinsics.  }
\end{figure}

Special-case processing is required when our arrays do not have lengths
divisible by eight: we need to finish the algorithm with a
small scalar intersection algorithm.
Special processing is also required if an array starts with zero,
because the \texttt{pcmpistrm} instruction is designed for string
values and it processes null values as special (string ending) values.

Evidently, we can easily modify this function if we only wish to get the
cardinality. For the most part, we merely skip the memory store instructions.

When the \texttt{pcmpistrm} instruction is available, as it is on all recent x64 processors, it is advantageous to replace a conventional list intersection function by such a vectorized approach. However, we still rely on a galloping intersection when one of the two input arrays is much smaller. For example, the vectorized approach is obviously not helpful to intersect an array made of 1000~values with one made of two~values. There are vectorized versions of the galloping intersection~\cite{simdcompandinter} which could further improve our performance in these cases, but we leave such an optimization for future work.

\subsection{Vectorized Unions  Between Arrays}
\label{sec:simdunion}
To compute the union between two array containers, we adapt an
approach originally developed for merge sort using SIMD
instructions~\cite{Inoue:2015:SCA:2809974.2809988}.  Given two sorted
vectors, we can generate two sorted output vectors that are the result
of a ``merger'' between the two input vectors, by using a sorting
network~\cite{Batcher:1968:SNA:1468075.1468121,Knuth:1998:ACP:280635}
whose branch-free implementation uses SIMD
minimum and maximum instructions.  We can compute
such a merger with eight minimum instructions and eight maximum
instructions, given two vectors made of eight 16-bit integers (see Fig.~\ref{code:simdmerge}).

We can put this fast merge function to good use by initially loading one vector from each sorted array and merging them,
so that the small values go into one vector ($B_1$) and the large values go into the other ($B_2$).
Vector $B_1$ can be queued for output.
We then advance a pointer in one of the two arrays, choosing the array with a smallest new value, and then load a new vector. We merge this new vector with $B_2$. 
We repeat this process; see Algorithm~\ref{algo:genunionalgo}.
 We terminate with a scalar merge because the 
arrays' lengths might not be a multiple of the vector size.

\begin{algorithm}
\begin{algorithmic}[1]
\REQUIRE two non-empty arrays made of a multiple of $K$ distinct values in sorted order
\STATE load a block $B_1$ of $K$ elements from the first array
\STATE load a block $B_2$ of $K$ elements from the second array
\STATE Merge blocks $B_1$ and $B_2$ so that the smallest elements are in block $B_1$ and the largest are in block $B_2$. Output the elements from $B_1$, after removing the duplicates
\WHILE {there is more data in both lists}
\IF{the next value loaded from the first array would be smaller than the next value in the second array}
\STATE load a new block $B_1$ of $K$ elements from the first array
\ELSE
\STATE load a new block $B_1$ of $K$ elements from the second array
\ENDIF
\STATE Merge blocks $B_1$ and $B_2$ so that the smallest elements are in block $B_1$ and the largest are in block $B_2$
\STATE  output the elements from $B_1$, after removing the duplicates while taking into account the last value written to output
\ENDWHILE
\WHILE {there is more data in one list}
\STATE load a new block $B_1$ of $K$ elements from the remaining array
\STATE Merge blocks $B_1$ and $B_2$ so that the smallest elements are in block $B_1$ and the largest are in block $B_2$,
\STATE output the elements from $B_1$, after removing the duplicates while taking into account the last value written to output
\ENDWHILE
\STATE Output the elements from $B_2$, after removing the duplicates
\end{algorithmic}
\caption{Generic block-based union algorithm\label{algo:genunionalgo}.}
\end{algorithm}

The queued vector might include duplicate values. To
``deduplicate'' them we use a single vectorized comparison
between each new vector with a version of itself
offset back by one element (taking care to
handle the final previously stored value). For the mask resulting
from this comparison, we can quickly extract the integers that need
to be written, using a look-up table and an appropriate permutation.  See Fig.~\ref{code:storeunique}.

Our vectorized approach is implemented to
minimize mispredicted
branches.
 We use 128-bit SIMD vectors instead of 256-bit AVX vectors. Larger vectors would further reduce the number of branches, but informal experiments indicate that the gains are modest. A downside of using larger vectors is that more of the processing occurs using scalar code, when the array lengths are not divisible by the vector lengths.

\begin{figure}[tbh]
\begin{center}
\begin{tabular}{c}
\begin{lstlisting}
void sse_merge(const __m128i *vInput1,
      const __m128i *vInput2,// input 1 & 2
      __m128i *vecMin, __m128i *vecMax) {
  vecTmp = _mm_min_epu16(*vInput1, *vInput2);
  *vecMax = _mm_max_epu16(*vInput1, *vInput2);
  vecTmp = _mm_alignr_epi8(vecTmp, vecTmp, 2);
  *vecMin = _mm_min_epu16(vecTmp, *vecMax);
  *vecMax = _mm_max_epu16(vecTmp, *vecMax);
  vecTmp = _mm_alignr_epi8(*vecMin, *vecMin, 2);
  *vecMin = _mm_min_epu16(vecTmp, *vecMax);
  *vecMax = _mm_max_epu16(vecTmp, *vecMax);
  vecTmp = _mm_alignr_epi8(*vecMin, *vecMin, 2);
  *vecMin = _mm_min_epu16(vecTmp, *vecMax);
  *vecMax = _mm_max_epu16(vecTmp, *vecMax);
  vecTmp = _mm_alignr_epi8(*vecMin, *vecMin, 2);
  *vecMin = _mm_min_epu16(vecTmp, *vecMax);
  *vecMax = _mm_max_epu16(vecTmp, *vecMax);
  vecTmp = _mm_alignr_epi8(*vecMin, *vecMin, 2);
  *vecMin = _mm_min_epu16(vecTmp, *vecMax);
  *vecMax = _mm_max_epu16(vecTmp, *vecMax);
  vecTmp = _mm_alignr_epi8(*vecMin, *vecMin, 2);
  *vecMin = _mm_min_epu16(vecTmp, *vecMax);
  *vecMax = _mm_max_epu16(vecTmp, *vecMax);
  vecTmp = _mm_alignr_epi8(*vecMin, *vecMin, 2);
  *vecMin = _mm_min_epu16(vecTmp, *vecMax);
  *vecMax = _mm_max_epu16(vecTmp, *vecMax);
  *vecMin = _mm_alignr_epi8(*vecMin, *vecMin, 2);
}
\end{lstlisting}
\end{tabular}
\end{center}

\caption{\label{code:simdmerge}Fast merge function using Intel SIMD instrinsics, takes two sorted arrays of eight 16-bit integers and produces two vectors (\texttt{vecMin}, \texttt{vecMax}) containing the sixteen integer inputs in a sorted sequence, with the eight smallest integers in \texttt{vecMin}, and eight largest in \texttt{vecMax}.  }
\end{figure}

\begin{figure}[tbh]
\begin{center}
\begin{tabular}{c}
\begin{lstlisting}[basicstyle=\small]
// returns how many unique values are found in n
// the last integer of "o" is last written value
int store_unique(__m128i o, __m128i n, uint16_t *output) {
    // new vector v starting with last value of "o"
    // and containing the 7 first values of "n"
    __m128i v = _mm_alignr_epi8(n, o, 16 - 2);
    // compare v and n and create a mask
    int M = _mm_movemask_epi8(
        _mm_packs_epi16(_mm_cmpeq_epi16(v,n), _mm_setzero_si128()));
    // "number" represents the number of unique values
    int number = 8 - _mm_popcnt_u32(M);
    // load a "shuffling" mask
    __m128i key = _mm_loadu_si128(uniqshuf + M);
    // the first "number" values of "val" are the unique values
    __m128i val = _mm_shuffle_epi8(n, key);
    // we store the result to "output"
    _mm_storeu_si128((__m128i *)output, val);
    return number;
}
\end{lstlisting}
\end{tabular}
\end{center}

\caption{\label{code:storeunique}Fast function that takes an vector containing sorted values with duplicates, and stores the values with the duplicates removed, returning how many values were written. We provide a second vector containing the last value written as its last component.
}
\end{figure}

\subsection{Vectorized Differences  Between Arrays}
\label{sec:simddifference}

Unlike union and intersection,
the difference operator is an asymmetric operation: $A	\setminus B$ differs from $B	\setminus A$. We remove the elements of the second array from the first array.

If we can compute the intersection quickly between blocks of data (e.g., with the \texttt{_mm_cmpistrm} intrinsic), 
we can also quickly obtain
the difference.  In our vectorized intersection algorithm (\S~\ref{sec:simdinter}), we identify the elements that are part of the intersection using a bitset. In a difference algorithm, picking blocks of values from the first array, we can successively remove the elements that are present in the second array, by iterating through its data in blocks. The intersection, expressed as a bitset, marks the elements for deletion, and we can conveniently accumulate these indications
 by taking the bitwise OR of the bitsets.
 Indeed, suppose that we want to compute  $A	\setminus B$ where $A= \{1,3,5,7\}$ and $B=\{1,2,3,4,6,7,8,9\}$ using blocks of 4~integers:
 \begin{itemize}
 \item We compute the intersection between the first (and only) block from $A$ given by $\{1,3,5,7\}$ and the first block from $B$ given by $\{1,2,3,4\}$. We find that $1,3$ are part of the intersection, but not $5,7$. We can represent this result with the bitset \texttt{0b0011} (or \texttt{0x3}).
 \item We advance to the next block in the second array, loading $\{6,7,8,9\}$. The intersection between the two blocks is $\{1,3,5,7\}\cap\{6,7,8,9\}=\{7\} $ which we can represent with the bitset  \texttt{0b1000}.
 \item We can compute the bitwise OR between the two bitsets \texttt{0b0011} and \texttt{0b1000} to get \texttt{0b1011}. It indicates that all but the third value in the block $\{1,3,5,7\}$ must be removed. So we output $\{3\}$.
 \end{itemize}
Algorithm~\ref{algo:gendiffalgo} describes the difference algorithm at a high level.

As with our other algorithms, we need to finish the computation using scalar code because the inputs may not be divisible by the vector size.
Moreover, if we only need to compute the size of the difference, we can use the same algorithm, but we simply do not store the result, only tracking the cardinality.

\begin{algorithm}
\begin{algorithmic}[1]
\REQUIRE two non-empty arrays made of a multiple of $K$ distinct values in sorted order
\STATE load a block $B_1$ of $K$ elements from the first array
\STATE load a block $B_2$ of $K$ elements from the second array
\WHILE {true}
\STATE mark for deletion the elements in $B_1\cap B_2 $
\IF{ $\max B_1 =  \max B_2$}
\STATE write $B_1$ to the output after removing the elements marked for deletion
\STATE load a new block $B_1$ of $K$ elements from the first array, or terminate if  exhausted
\STATE load a new block $B_2$ of $K$ elements from the second array, or break the loop if exhausted
\ELSIF{ $\max B_1 <  \max B_2$}
\STATE write $B_1$ to the output  after removing the elements marked for deletion
\STATE load a new block $B_1$ of $K$ elements from the first array, or terminate if exhausted
\ELSIF{ $\max B_2 <  \max B_1$}
\STATE load a new block $B_2$ of $K$ elements from the second array, or break the loop if exhausted
\ENDIF
\ENDWHILE
\STATE write $B_1$ after removing the elements marked for deletion as well as all unprocessed data from the first array  to the output
\end{algorithmic}
\caption{Generic block-based difference algorithm\label{algo:gendiffalgo}.}
\end{algorithm}

\subsection{Vectorized Symmetric Differences Between Arrays}
\label{sec:simdsymmetricdifference}

The symmetric difference can be computed using a block-based algorithm similar to the union algorithm (Algorithm~\ref{algo:genunionalgo}). When computing the union, we remove duplicates, keeping just one instance of a value repeated twice. E.g., given $\{1,1,2,2,3\}$, we want to output $\{1,2,3\}$. When computing the symmetric difference, we need to entirely remove a value if it is repeated. Thus, given $\{1,1,2,2,3\}$, we want to get the set $\{3\}$.
We can use a fast vectorized function for this purpose (see Fig.~\ref{code:storeuniquexor}), and it is similar to the corresponding deduplication function used for the vectorized union (Fig.~\ref{code:storeunique}).

In the union, we can immediately write all values after removing duplicates. With the symmetric difference, the largest value in the current block should not be written immediately. Indeed, we need to account for the fact that the next block might start with this largest value, and that it may become necessary to omit it. Thus, we never write the current largest value of the current block when computing the symmetric difference.
%
Instead, it is (potentially) added when the next block is processed --- note that \texttt{val} in Fig.~\ref{code:storeuniquexor} is computed
from \texttt{v2} (which has this value), rather than from \texttt{n}.
Of course, when the algorithm terminates, we must write the largest value encountered if it were not repeated. In any case, we need to terminate the algorithm with scalar code given that the size of  our input lists may not be divisible by the vector size.
%
%
%
%
%

\begin{figure}[tbh]
\begin{center}
\begin{tabular}{c}
\begin{lstlisting}[basicstyle=\small]
int store_xor(__m128i o, __m128i n, uint16_t *output) {
    __m128i v1 = _mm_alignr_epi8(n, o, 16 - 4);
    __m128i v2 = _mm_alignr_epi8(n, o, 16 - 2);
    __m128i el = _mm_cmpeq_epi16(v1, v2);
    __m128i er = _mm_cmpeq_epi16(v2, n);
    __m128i erl = _mm_or_si128(el,er);
    int M = _mm_movemask_epi8(
        _mm_packs_epi16(erl, _mm_setzero_si128()));
    int nv = 8 - _mm_popcnt_u32(M);
    __m128i key = _mm_lddqu_si128(uniqshuf + M);
    __m128i val = _mm_shuffle_epi8(v2, key);
    _mm_storeu_si128((__m128i *)output, val);
    return nv;
}
\end{lstlisting}
\end{tabular}
\end{center}

\caption{\label{code:storeuniquexor}Fast function taking a vector containing sorted values with duplicates and entirely removing the duplicated values, returning how many values were written. We provide a second vector containing the last value written as its last component. }
\end{figure}

\section{Experiments}
\label{sec:exp}
We present  experiments
on realistic datasets. Our benchmarks 
directly carry out operations
on collections of in-memory sets. These original sets are immutable: they are not modified during the benchmarks.  For a validation of Roaring in a database system with external memory processing, see for example Chambi et al.~\cite{Chambi:2016:ODR:2938503.2938515} where they showed that switching to Roaring from Concise improved system-level benchmarks.

\subsection{Software}
\label{sec:software}

We use the GNU GCC compiler (version 5.4) on a Linux server. All code was compiled with full optimization (\texttt{-O3 -march=native}) as it provided the best performance.

We published our C code as the CRoaring library; it is available on GitHub under an Apache license\footnote{\url{https://github.com/RoaringBitmap/CRoaring}}. The CRoaring library is amalgamated into a single source file (\texttt{roaring.c}) and a single header file (\texttt{roaring.h})~\cite{sqliteamalgamation}.
The code is portable, having been tested on several operating systems (Linux, Windows, macOS), on several processors (ARM, x64) and on several compilers (GNU GCC, LLVM's Clang, Intel, Microsoft Visual Studio). We offer interfaces to the library in several other languages (e.g., C\#, C++, Rust, Go, Python); it is also available as a module of the popular open-source in-memory database Redis~\cite{redis}.
When instantiating the original Roaring bitmaps, we call the functions \texttt{roaring_bitmap_run_optimize} and \texttt{roaring_bitmap_shrink_to_fit} to minimize memory usage.
We use version~0.2.33 of the CRoaring library in our experiments.

To assess the library and compare it with alternatives, we prepared benchmarking software in C and C++, also freely available on GitHub.\footnote{\url{https://github.com/RoaringBitmap/CBitmapCompetition}} We consider the following alternatives:
\begin{itemize}
\item We use the C++ BitMagic library\footnote{\url{https://sourceforge.net/projects/bmagic/}} (version 3.7.1, published in July 2016). During compilation, we enable the SIMD optimizations offered by the library by defining the appropriate flags (\texttt{BMSSE2OPT} and \texttt{BMSSE42OPT}). We instantiate bitmaps using the default C++ template (\texttt{bm::bvector<>}), since there is no other documented alternative. We follow the  documented approach to optimizing memory usage, calling \texttt{set\_new\_blocks_strat(bm::BM_GAP)} followed by
\texttt{optimize()} after creating the original bitmaps. We do not call \texttt{optimize()} during the timed benchmarks as it may affect negatively the timings.

During the optimization procedure, BitMagic allocates a temporary  \SI{8}{kB}~memory block that fails to get deallocated. This adds an unnecessary \SI{8}{kB} to the memory usage of each bitmap, an undesirable effect. Because this memory allocation is part of the private code of the library, we cannot manually deallocate it without modifying the library. However, as a workaround, we copy each of the original bitmaps after calling \texttt{optimize()}  using the copy constructor: this new copy does not have such a useless \SI{8}{kB}~memory block.

This extra copy  to improve BitMagic's memory usage makes the creation of the bitmaps slightly slower, but it does not harm  BitMagic's timings in our benchmarks.
\item We use the C++ EWAHBoolArray library for comparison with the EWAH format (version 0.5.15).\footnote{\url{https://github.com/lemire/EWAHBoolArray}}
Code from this library is part of the  Git version-control tool~\cite{gitewah}. After creating the original bitmaps, we call their \texttt{trim} method to minimize memory usage. The \texttt{trim} method is not called during the timed benchmarks.
\item For comparison against uncompressed bitsets, we 
developed
 a portable C library, cbitset\footnote{\url{https://github.com/lemire/cbitset}} (version 0.1.6). The library is written using portable C, without processor-specific optimizations. It uses 64-bit words. To represent a set of integers in $[0,n)$, it attempts to use as little as $\lceil n/64 \rceil \times 8$~bytes, plus some overhead.
\item For comparison against the WAH and Concise formats~\cite{colantonio:2010:ccn:1824821.1824857}, we 
use
the Concise C++ library\footnote{\url{https://github.com/lemire/Concise}} (version 0.1.8).
Since Concise and WAH are similar formats, we can support both of them using the same C++ code without sacrificing performance with a C++ template. After creating the original bitmaps, we call their \texttt{compact} method to minimize memory usage, but never during the timed benchmarks.
\end{itemize}
The EWAH, WAH, and Concise formats can work with words of various sizes (e.g., 32~bits, 64~bits). When using 64 bits, we sometimes gain some speed at the expense of a significant increase in memory usage~\cite{lemire2016consistently}. For simplicity, we only consider 32-bit words, thus maximizing compression.

We also include the standard C++ library bundled with our compiler which provides \texttt{std::vector} and \texttt{std::unordered_set}.
\begin{itemize}
\item
 We use  \texttt{std::unordered_set} as the basis for testing hash-set performance.
The \texttt{std::unordered_set} template is hardly the sole way to implement a hash set in C++, but since it is part of the standard library, 
we consider it
a good default implementation. To reduce memory usage, we call
 the \texttt{rehash} method on our \texttt{std::unordered_set} instances after constructing them, passing the number of items as an argument.
Implementing set operations for \texttt{std::unordered_set} is straightforward. For example, to compute an intersection between two \texttt{std::unordered_set}, we pick the smallest, iterate over its content and check each value in the largest one.
Our implementations have good computational complexity.
Thus computing the intersection between two hash sets of size $n_1$ and $n_2$ has complexity $O(\min(n_1,n_2))$, the union and symmetric differences have complexity $O(n_1 + n_2)$, and the difference has complexity $O(n_1)$.
\item
To minimize the memory usage of our original \texttt{std::vector} instances, we call their \texttt{shrink_to_fit} method after constructing them, but never during the timed benchmarks.
To implement set operations over the \texttt{std::vector} data structure, we keep the values sorted and make use of the functions provided by STL (\texttt{std::set_intersection},  \texttt{std::set_union},
 \texttt{std::set_difference},
   \texttt{std::set_symmetric_difference}, \ldots).
These functions take iterators pointing at the inputs and output, so we can use them to, for example, measure the size of the output without necessarily materializing it.
   To locate a value within an \texttt{std::vector}, we use \texttt{std::binary_search}, a logarithmic-time function.
\end{itemize}

Of course, when comparing different software libraries, we expect performance differences due to specific implementation issues. Though this is unavoidable, we alleviate the problem by ensuring that all of our code, and all of the libraries' code is freely available, thus easing reproducibility and further work.

\subsection{Hardware}
\label{sec:hardware}
We use   a Linux server with an Intel i7-6700 processor  (Skylake microarchitecture, \SI{3.4}{GHz}, \SI{32}{kB} of L1 data cache, \SI{256}{kB} of L2 cache per core and \SI{8}{MB} of L3 cache). The server
 has \SI{32}{GB} of RAM (DDR4~2133, double-channel).
Our benchmarks 
do not access disk
and are single-threaded.
The processor always runs at its highest clock speed and Turbo Boost is disabled.
Times are collected using the \texttt{rdtsc} instruction, as described in Intel's documentation~\cite{intelbenchmark}.
Prior to benchmarking, we checked that all results could be reproduced reliably within a margin of error of \SI{5}{\percent} or better.

\subsection{Data}

We focus on the  cases where the integer values  span a wide range of values and where  sets contain many integers.
Thus, we make use of eight datasets  from earlier work~\cite{lemire2016consistently,LemireKaserGutarra-TODS}.
Our datasets are derived from bitmap indexes built on real-world tables.
Some are taken as-is while others are generated from tables
that have been
lexicographically sorted prior to indexing~\cite{arxiv:0901.3751}. We have two sets of bitmaps
from each data source (e.g., \Censeighten{} and \Censeightensrt{}).
For each dataset, we randomly chose 200~sets, each of which contains the record ids of
rows matching predicate $A=v$, for some column $A$ and column-value $v$.
 We present the basic characteristics
of these sets in Table~\ref{tab:chrac}. All of the datasets are publicly available
as part of our CRoaring software library. The universe size of a dataset is the smallest value $n$
such that all sets are contained in  $[0,n)$. The density is the cardinality divided by the universe size.

\begin{table}[bh]
\caption{\label{tab:chrac}Characteristics of
our realistic datasets}
\centering
\begin{tabular}{c|rrr}
\toprule
 & universe  & average cardi- & average \\
&  size & nality per set & density \\
        \midrule
\CensInc   & \num{199523} & \num{34610.1} & \num{0.17} \\
\CensIncsrt   & \num{199523} & \num{30464.3} & \num{0.15} \\
\Censeighten   & \num{4277806} & \num{5019.3} & \num{0.001} \\
\Censeightensrt   & \num{4277735} & \num{3404.0} & \num{0.0008}\\
\Weather   & \num{1015367} & \num{64353.1} & \num{0.063} \\
\Weathersrt  & \num{1015367} & \num{80540.5} & \num{0.079}\\
\Wikileaks  & \num{1353179} & \num{1376.8} & \num{0.001} \\
\Wikileakssrt  & \num{1353133} & \num{1440.1} & \num{0.001}\\
\bottomrule
\end{tabular}
\end{table}

We take our datasets as reasonable examples that can serve to
illustrate the benefits (as well as the lack of benefit) of our Roaring
optimizations. We stress that no single data structure (bitset, array,
hash set, compressed bitset) can be best for all datasets and all
applications. The choice depends on the data  characteristics and is non-trivial~\cite{insights2017}.

We want to benchmark computational efficiency to evaluate our optimization. Thus we want to avoid disk access entirely.
Hence our datasets easily fit in
memory by design. We further validate our results with a large in-memory dataset in Appendix~\ref{appendix:big}.

In our benchmarks, the original collection of sets is immutable. For example, if we compute the intersection between two sets, we generate a new data structure with the result. This reflects a common usage scenario: e.g., Druid~\cite{Chambi:2016:ODR:2938503.2938515} relies on collections of immutable and persistent compressed bitmaps to accelerate queries.

\subsection{Memory Usage}

Data structures that fail to fit in memory must be retrieved from slower storage media. Thus, everything else being equal, we want our data structures to use as little memory as possible.

Exactly computing the memory usage of complex data structures in C and C++ can be difficult; the same C code running with different libraries or a different operating system might use memory differently. Thus we need to proceed with care.

For the \texttt{std::vector} implementation, we know from first principles that the memory usage (when ignoring a small amount of constant overhead) is 32~bits per entry. For the  hash sets (\texttt{std::unordered_set}), we use a custom memory allocator that keeps track of allocation requests and record the provided byte counts, assuming for simplicity that memory allocations carry no overhead.   For continuous storage, this estimate should be accurate, but for a hash set (where there are numerous allocations), we underestimate the true memory usage. Because our custom memory allocation might add an unnecessary computational overhead, we run the timing tests without it. 

We proceed similarly with Roaring. Because we implemented Roaring using the C language, we instrument 
the malloc/free calls so that they register the memory allocated and freed. We also run our experiments with and without this instrumentation, to make sure that we can record the speed without undue overhead. The CRoaring library supports a compact and portable serialization format.
 For most of our datasets, the in-memory size and the compact serialized size  are nearly identical. There are only three datasets where there were noteworthy differences between in-memory and serialized sizes: \Censeightensrt{} (2.77~bits per value vs. 2.16~bits per value),
\Wikileaks{} (2.58~bits per value vs. 1.63~bits per value),
  and
\Wikileakssrt{} (7.04~bits per value vs.  5.89~bits per value).
Not coincidentally, these are our datasets with the fewest values per set. The difference can be explained by the overhead introduced by each new container. It suggests that the memory layout in the CRoaring library could be improved in some cases, by reducing the memory overhead of the containers.

For uncompressed bitsets, EWAH, WAH, and Concise, we rely on the memory usage in bytes reported by the corresponding software libraries. Because they store all their data in a single array, we can be confident that this approach 
 provides a reliable count.

For the BitMagic library, we rely on the numbers provided by the \texttt{calc_stat} method, retrieving the \texttt{memory_used} attribute. An examination of the source code reveals that it should provide a fair estimate of the memory usage.

\begin{table*}
\caption{\label{tab:memusage}Memory usage in bits per value. For each dataset, the
best result is in bold. The hash set is implemented using the STL's (\texttt{unordered_set}).}
\centering\small
\begin{tabular}{c|rrrr>{\columncolor[gray]{0.9}}rrrr}
\toprule
 & bitset& \texttt{vector}  & hash set & BitMagic & Roaring & EWAH & WAH & Concise\\
        \midrule
\CensInc{}  & 5.66 & 32.0 & 195 & 4.46 & \textbf{2.60} & 3.29 & 3.37 & 2.94 \\
\CensIncsrt{}  & 6.01 & 32.0 & 195 & 1.89 & 0.600 & 0.640 & 0.650 & \textbf{0.550} \\
\Censeighten{}  & 524 & 32.0 & 195 & 46.3 & \textbf{15.1} & 33.8 & 34.3 & 25.6 \\
\Censeightensrt{}  & 888 & 32.0 & 195 & 16.0 & \textbf{2.16} & 2.91 & 2.95 & 2.48 \\
\Weather{}  & 15.3 & 32.0 & 195 & 8.79 & \textbf{5.38} & 6.67 & 6.82 & 5.88 \\
\Weathersrt{}  & 11.4 & 32.0 & 195 & 0.960 & \textbf{0.340} & 0.540 & 0.540 & 0.430 \\
\Wikileaks{}  & 796 & 32.0 & 195 & 29.8 & \textbf{5.89} & 10.8 & 10.9 & 10.2 \\
\Wikileakssrt{}  & 648 & 32.0 & 195 & 24.0 & \textbf{1.63} & 2.63 & 2.67 & 2.23 \\

\bottomrule
\end{tabular}
\end{table*}

We present the estimated memory usage in Table~\ref{tab:memusage}:

\begin{itemize}
\item  The memory usage of the uncompressed bitset approach can be excessive (up to 888~bits per 32-bit value). However, it is less than 32~bits per value in four datasets: \CensInc{}, \CensIncsrt{}, \Weather{} and \Weathersrt{}.
\item The chosen hash-set implementation is clearly not economical memory-wise, using 195~bits per 32-bit integer. The underlying data structure is a hash table composed of an array of buckets. Because the maximum load factor is 1.0 by default, we have at least as many buckets as we have values. Each bucket is a singly linked list starting with a 64-bit pointer to the first value node. Each value node uses 128~bits: 64~bits for a pointer to the next value node, and 64~bits to store the actual integer value. Though we use 32-bit integers as values, the implementation pads the node size to 128~bits for for memory alignment. Thus at least 192~bits ($64+128$) are necessary to store each value, which explains the 195~bits encountered in practice in our datasets.
\item The memory usage of EWAH, WAH, and Concise are similar
to one another (within 20\%).  Concise is slightly more economical in most cases. Roaring uses less memory than Concise in all datasets except \CensIncsrt{}, where Concise uses 10\% less memory. In two cases (\Censeighten{} and \Wikileaks{}), Roaring uses much less memory than Concise (30--40\%).
\item  In some cases, BitMagic uses a surprising amount of memory (up to 46.3~bits per 32-bit integer). Something similar was observed by Pieterse et al.~\cite{Pieterse:2010:PCB:1899503.1899530} who wrote that ``although it [BitMagic] maintained acceptable performance times, its
memory usage is exorbitant.''  The BitMagic library does not appear to be optimized for low memory usage.
\end{itemize}

\subsection{Sequential Access}

It is useful to be able to iterate through all values of a set. All of the data structures under consideration allow for efficient in-order access, except for the hash set. The fastest data structure for this problem is the sorted array. Setting it aside, we compare the other software implementations in Table~\ref{tab:iteration}. In this benchmark, we simply go through all values, as a means to compute the total cardinality. WAH and Concise have similar performance, with WAH being marginally faster. EWAH has also similar performance, being slightly faster for some datasets, and slightly slower for others. Depending on the dataset, the worst performance is either with the  bitset or the hash set. Indeed, iterating through a bitset can be slow when it is memory inefficient.

The hash set, BitMagic, EWAH, WAH and Concise implementation rely on an iterator approach, as is typical in C++. Our benchmark code takes the form of a simple for loop.
\begin{lstlisting}
for(auto j = b.begin(); j != b.end(); ++j)
  count++;
\end{lstlisting}

The bitset and CRoaring implementations favor an approach that is reminiscent of the functional for-each construct, as implemented by STL's \texttt{std::for_each} and Java's \texttt{forEach}.
They take a function as a parameter, calling the function for each element of the set, continuing as long as the function returns true. The expected function signature has an optional void-pointer parameter.
\begin{lstlisting}
typedef bool (*roaring_iterator)(uint32_t value, void *param);
\end{lstlisting}
In our benchmark, the void pointer is used to pass a variable that gets incremented with each value encountered.

\begin{table*}
\caption{\label{tab:iteration}Time needed to iterate through all values, checking the cardinality in the process. We report the number of CPU cycles used per value.}
\centering\small
\begin{tabular}{c|rrr>{\columncolor[gray]{0.9}}rrrr}
\toprule
 & bitset  & hash set & BitMagic & Roaring & EWAH & WAH & Concise\\
        \midrule
\CensInc{}                 & 7.53                  & 28.5              & 20.4              & \textbf{5.87}              & 13.1              & 9.23              & 9.34              \\
\CensIncsrt{}              & 6.79                   & 12.6              & 6.86              & \textbf{5.33}              & 6.03              & 7.63              & 7.85              \\
\Censeighten{}             & 42.7                   & 67.5              & 70.9              & \textbf{5.16}              & 29.4              & 29.4              & 31.7              \\
\Censeightensrt{}          & 41.4                  & 20.4              & 11.0              & \textbf{6.32}              & 9.10              & 10.0              & 10.1              \\
\Weather{}                 & 10.0                   & 40.7              & 31.7              & \textbf{6.43}              & 15.9              & 12.4              & 12.8              \\
\Weathersrt{}              & 6.73                   & 12.9              & 6.79              & \textbf{5.25}              & 5.87              & 7.46              & 7.68              \\
\Wikileaks{}               & 46.2                 & 24.6              & 12.2              & \textbf{9.61}              & 18.4              & 16.1              & 16.6              \\
\Wikileakssrt{}            & 33.2              & 13.1              & 7.96              & \textbf{5.86}              & 7.95              & 8.95              & 9.19              \\
\bottomrule
\end{tabular}
\end{table*}

\subsection{Membership Queries}

All of the software libraries under consideration support membership queries: given an integer $x$, we can ask whether it belongs to the set. Recall that for the \texttt{std::vector} tests, we use the \texttt{std::binary_search} function provided by the standard library.

To test the performance of random access, for each data set, we compute the size of the universe ($[0,n)$) and check for the presence of the integers $\lfloor n/4 \rfloor$,  $\lfloor n/2 \rfloor$ and $\lfloor 3 n/4 \rfloor$ in each set of the dataset.
We present the results in Table~\ref{tab:randaccess}.

The bitset consistently has the best performance, using less than a handful of CPU cycles per query.  The hash set and BitMagic have the next best random-access performance, followed by Roaring, followed by the arrays (\texttt{vector}). The arrays  have much worse performance than Roaring, sometimes by an order of magnitude.

Roaring is inferior to BitMagic in these tests even though they implement a similar two-level data structure. Two design elements can explain this inferior performance:
\begin{itemize}
\item Roaring uses a binary search is used to first locate (if it exists) a corresponding container, before searching the container. BitMagic does not require such a binary search because, given values in the range $[0,n)$, it materializes $\lceil n/2^{16} \rceil$~containers in a flat array, with containers marked as empty or full as needed. Thus BitMagic can locate the corresponding array directly, without any branching.
\item Roaring can store up to \num{4096}~values in array containers  whereas BitMagic never stores more than \num{1280}~values in its array containers. So where Roaring uses large array containers, BitMagic use bitset containers. It is  faster to look up the value of a bit within a bitset container than to do a binary search within an array container having more than 1280~values.
\end{itemize}
Both of these elements also contribute to BitMagic's higher memory usage.

Both the bitset and the hash set have expected constant-time random-access complexity. However, the bitset is an order of magnitude faster in this instance.


EWAH, WAH, and Concise have terrible random-access performance as expected, with EWAH faring slightly better. Indeed, unlike our other data structures, these suffer from a linear-time random-access complexity, meaning that the time required to check the value of a bit is proportional to the compressed size of the bitmap. Though Concise usually compresses better than either EWAH or WAH, we see that it can also have a slightly worse random access performance.

\begin{table*}[h]
\caption{\label{tab:randaccess}Performance comparisons for random-access checks (membership queries). We report the number of CPU cycles used on average per queries.}
\centering\small
\begin{tabular}{c|rrrr>{\columncolor[gray]{0.9}}rrrr}
\toprule
 & bitset& \texttt{vector}  & hash set & BitMagic & Roaring & EWAH & WAH & Concise\\
        \midrule
\CensInc{}  & \textbf{3.74} & 415 & 49.2 & 33.6 & 63.6 & 3260 & 19300 & 19100 \\
\CensIncsrt{}  & \textbf{3.70} & 547 & 46.7 & 49.7 & 67.4 & 947 & 3690 & 3800 \\
\Censeighten{}  & \textbf{2.36} & 174 & 38.2 & 11.6 & 17.4 & 8820 & 26700 & 30000 \\
\Censeightensrt{}  & \textbf{2.51} & 159 & 36.6 & 11.0 & 18.3 & 561 & 1810 & 1880 \\
\Weather{}  & \textbf{3.54} & 564 & 53.0 & 27.4 & 76.8 & 13700 & 65200 & 69000 \\
\Weathersrt{}  & \textbf{3.47} & 661 & 48.8 & 42.0 & 53.7 & 2270 & 7490 & 7330 \\
\Wikileaks{}  & \textbf{3.68} & 172 & 45.4 & 20.9 & 27.7 & 1030 & 2410 & 2350 \\
\Wikileakssrt{}  & \textbf{2.79} & 152 & 43.2 & 14.8 & 20.1 & 248 & 531 & 493 \\
\bottomrule
\end{tabular}
\end{table*}

\subsection{Intersections, Unions, Differences, and Symmetric Differences}
\label{sec:wholestuff}

Recall that our datasets are made of 200~sets each.
We consider the 199~intersections, unions, differences, and symmetric differences  between successive sets. Our functions leave the inputs unchanged, but they materialize the result as a new set 
and so the
timings include the construction
of    
%
a new data structure. After each computation,  we check
the cardinality of the result against a precomputed result---thus helping to  prevent  the compiler from optimizing  the computation.  See Table~\ref{tab:perf}.

The relative rankings of the different implementations is similar, whether we consider intersections, unions, differences, or symmetric differences. Intersections run faster, whereas unions and symmetric differences take longer.

What is immediately apparent is that for the \CensInc{}, \CensIncsrt{}, and \Weather{} datasets, an uncompressed bitset is faster. However, Roaring comes in second, having sometimes half the speed. However, Roaring uses 10$\times$ less memory than an uncompressed bitset on \CensIncsrt{}. The performance of an uncompressed bitset is much worse on several other datasets such as \Censeighten{}, \Censeightensrt{}, \Wikileakssrt{} and \Wikileaks{}.

Excluding the uncompressed bitset, the fastest data structure is generally Roaring. On the \Censeighten{} dataset, Roaring is an order of magnitude faster than any other approach at computing intersections, and generally much faster.

In most instances, the hash set offers the worst performance. This may seem surprising as hash sets have good computational complexity. However, hash sets have also poor locality: i.e., nearby values are not necessarily located in nearby cache lines.

%

 \begin{table*}
\caption{\label{tab:perf}Performance comparisons for intersections, unions, differences and symmetric differences of integer sets. We report the number of CPU cycles used, divided by the number of input values. For each dataset, the best result is in bold.}
\centering

\subfloat[\label{tab:intersection}Two-by-two intersections: given 200~sets, we compute 199~intersections  and check the cardinality of each result.]{%
\centering\small
\begin{tabular}{c|rrrr>{\columncolor[gray]{0.9}}rrrr}
\toprule
 & bitset& \texttt{vector}  & hash set & BitMagic & Roaring & EWAH & WAH & Concise\\
        \midrule
\CensInc{}                 & \textbf{0.130}    & 7.84              & 55.3              & 0.380             & 0.220             & 1.41              & 1.80              & 2.15              \\
\CensIncsrt{}              & \textbf{0.150}    & 4.52              & 28.5              & 0.250             & 0.160             & 0.280             & 0.450             & 0.560             \\
\Censeighten{}             & 19.5              & 3.28              & 0.480             & 1.07              & \textbf{0.090}    & 3.74              & 14.5              & 20.9              \\
\Censeightensrt{}          & 33.0              & 3.96              & 1.41              & 0.830             & \textbf{0.140}    & 0.500             & 1.62              & 1.94              \\
\Weather{}                 & 0.420             & 7.36              & 35.5              & 0.590             & \textbf{0.380}    & 2.64              & 3.58              & 4.53              \\
\Weathersrt{}              & 0.330             & 4.36              & 15.1              & 0.160             & \textbf{0.080}    & 0.190             & 0.340             & 0.410             \\
\Wikileaks{}               & 26.4              & 4.15              & 14.9              & 3.06              & \textbf{1.43}     & 2.86              & 6.10              & 6.38              \\
\Wikileakssrt{}            & 19.6              & 4.15              & 26.5              & 1.92              & \textbf{0.580}    & 0.730             & 1.28              & 1.43              \\
\bottomrule
\end{tabular}
}\\
\subfloat[\label{tab:union}Two-by-two unions: given 200~sets, we compute 199~unions and check the cardinality of each result.]{%
\centering\small
\begin{tabular}{c|rrrr>{\columncolor[gray]{0.9}}rrrr}
\toprule
 & bitset& \texttt{vector}  & hash set & BitMagic & Roaring & EWAH & WAH & Concise\\
        \midrule
\CensInc{}                 & \textbf{0.130}    & 7.69              & 210               & 0.350             & 0.270             & 2.02              & 2.11              & 2.44              \\
\CensIncsrt{}              & \textbf{0.140}    & 5.45              & 173               & 0.270             & 0.300             & 0.560             & 0.650             & 0.760             \\
\Censeighten{}             & 13.6              & 6.10              & 265               & 3.68              & \textbf{1.13}     & 28.2              & 29.3              & 33.1              \\
\Censeightensrt{}          & 23.7              & 6.18              & 201               & 2.47              & \textbf{0.950}    & 2.50              & 3.03              & 3.38              \\
\Weather{}                 & \textbf{0.400}    & 7.70              & 250               & 0.600             & 0.570             & 4.46              & 4.59              & 5.41              \\
\Weathersrt{}              & 0.310             & 6.13              & 181               & 0.200             & \textbf{0.170}    & 0.480             & 0.560             & 0.620             \\
\Wikileaks{}               & 20.6              & 6.95              & 229               & 5.74              & \textbf{3.87}     & 9.23              & 10.9              & 11.3              \\
\Wikileakssrt{}            & 18.4              & 7.15              & 182               & 4.20              & \textbf{2.16}     & 2.37              & 2.57              & 2.63              \\
\bottomrule
\end{tabular}
}
\\
\subfloat[\label{tab:difference}Two-by-two differences: given 200~sets, we compute 199~differences and check the cardinality of each result.]{%
\centering\small
\begin{tabular}{c|rrrr>{\columncolor[gray]{0.9}}rrrr}
\toprule
 & bitset& \texttt{vector}  & hash set & BitMagic & Roaring & EWAH & WAH & Concise\\
        \midrule
\CensInc{}                 & \textbf{0.120}    & 8.38              & 177               & 0.340             & 0.280             & 1.72              & 1.98              & 2.34              \\
\CensIncsrt{}              & \textbf{0.120}    & 5.54              & 134               & 0.260             & 0.240             & 0.420             & 0.540             & 0.650             \\
\Censeighten{}             & 11.1              & 5.03              & 236               & 2.56              & \textbf{0.610}    & 15.8              & 21.9              & 26.7              \\
\Censeightensrt{}          & 19.4              & 5.76              & 171               & 2.94              & \textbf{0.570}    & 1.46              & 2.31              & 2.59              \\
\Weather{}                 & \textbf{0.390}    & 8.32              & 248               & 0.550             & 0.440             & 3.60              & 4.28              & 5.10              \\
\Weathersrt{}              & 0.290             & 5.78              & 149               & 0.210             & \textbf{0.150}    & 0.340             & 0.450             & 0.510             \\
\Wikileaks{}               & 18.4              & 6.05              & 215               & 6.00              & \textbf{2.65}     & 5.98              & 8.25              & 8.44              \\
\Wikileakssrt{}            & 15.2              & 5.74              & 161               & 4.43              & \textbf{1.27}     & 1.59              & 1.93              & 1.99              \\
\bottomrule
\end{tabular}
}
\\
\subfloat[\label{tab:symdiff}Two-by-two symmetric differences: given 200~sets, we compute 199~symmetric differences and check the cardinality.]{%
\centering\small
\begin{tabular}{c|rrrr>{\columncolor[gray]{0.9}}rrrr}
\toprule
 & bitset& \texttt{vector}  & hash set & BitMagic & Roaring & EWAH & WAH & Concise\\
        \midrule
\CensInc{}                 & \textbf{0.120}    & 8.79              & 266               & 0.350             & 0.370             & 1.96              & 2.04              & 2.38              \\
\CensIncsrt{}              & \textbf{0.130}    & 5.99              & 215               & 0.280             & 0.350             & 0.620             & 0.660             & 0.790             \\
\Censeighten{}             & 13.2              & 6.38              & 369               & 3.61              & \textbf{1.13}     & 27.5              & 28.9              & 33.3              \\
\Censeightensrt{}          & 22.5              & 6.41              & 243               & 2.56              & \textbf{0.950}    & 2.47              & 2.97              & 3.38              \\
\Weather{}                 & \textbf{0.400}    & 8.72              & 359               & 0.590             & 0.760             & 4.37              & 4.50              & 5.41              \\
\Weathersrt{}              & 0.310             & 6.23              & 237               & \textbf{0.220}    & \textbf{0.220}    & 0.510             & 0.560             & 0.630             \\
\Wikileaks{}               & 20.0              & 7.41              & 301               & 5.60              & \textbf{3.92}     & 8.96              & 10.7              & 11.4              \\
\Wikileakssrt{}            & 17.7              & 6.67              & 228               & 4.24              & \textbf{1.95}     & 2.33              & 2.51              & 2.66              \\
\bottomrule
\end{tabular}
}
\end{table*}

\subsection{Wide Unions}

In applications, it is not uncommon that we need to compute the union of many sets at once. Thus we benchmark the computation of the union of all 200~sets within a dataset.
Though various optimizations of this problem are possible,
 we  proceed sequentially, taking the first set, computing the union with the second set and so forth. This simple approach uses little memory and provides good performance.
In earlier work, this sequential approach was compared  to other strategies in some detail~\cite{lemire2016consistently}.

Though we leave the inputs unchanged, the bitset, BitMagic, and Roaring implementations can do some of the operations in-place. For example, we can copy the first bitset, and then compute the bitwise OR operation in-place with the following bitsets. Roaring has a convenience function to compute the union of several bitmaps at once (\texttt{roaring_bitmap_or_many}).

We present our result in  Table~\ref{tab:wideunion}.
Roaring is generally faster, except for two cases:  
\begin{itemize}
\item On \CensIncsrt{}, the uncompressed bitset is slightly faster (by about 20\%).
\item On \Censeighten{}, BitMagic is slightly faster than Roaring (by about 10\%).
\end{itemize}
In several instances, Roaring is several times faster than any other alternative.

 \begin{table*}
\caption{\label{tab:wideunion}Given 200~sets, we compute a single union,  we report the number of CPU cycles used, divided by the number of input values.}
\centering\small
\begin{tabular}{c|rrrr>{\columncolor[gray]{0.9}}rrrr}
\toprule
 & bitset& \texttt{vector}  & hash set & BitMagic & Roaring & EWAH & WAH & Concise\\
        \midrule
\CensInc{}                 & 0.090             & 43.4              & 98.4              & 0.160             & \textbf{0.050}    & 0.500             & 1.83              & 2.25              \\
\CensIncsrt{}              & \textbf{0.100}    & 37.2              & 60.2              & 0.190             & 0.110             & 0.160             & 0.470             & 0.590             \\
\Censeighten{}             & 9.85              & 542               & 1010              & \textbf{2.18}     & 2.56              & 276               & 388               & 402               \\
\Censeightensrt{}          & 16.2              & 619               & 353               & 4.58              & \textbf{3.26}     & 172               & 198               & 225               \\
\Weather{}                 & 0.350             & 94.1              & 237               & 0.280             & \textbf{0.160}    & 2.35              & 4.92              & 6.02              \\
\Weathersrt{}              & 0.250             & 69.1              & 82.3              & 0.210             & \textbf{0.040}    & 0.320             & 0.580             & 0.690             \\
\Wikileaks{}               & 15.4              & 515               & 413               & 11.9              & \textbf{3.01}     & 328               & 415               & 450               \\
\Wikileakssrt{}            & 14.5              & 496               & 274               & 8.92              & \textbf{1.94}     & 116               & 133               & 149               \\
\bottomrule
\end{tabular}
\end{table*}

\subsection{Fast Counts}

Sometimes, we do not wish the materialize the result of an intersection or of a union, we merely want to compute the
cardinality of the result as fast as possible.
For example, one might want to compute the Jaccard index (Tanimoto similarity) between two sets,($\vert A \cap B \vert  /\vert  A \cup B\vert $) or the cosine similarity ($\vert A \cap B \vert / \sqrt{\vert A\vert \vert B\vert }$).

Given a Roaring data structure, we can compute the total cardinality efficiently. For all bitset and array containers, we can simply look up the cardinality, whereas a simple sum over the length of all runs
can determine the cardinality of a run container. The same is true with other data structures such as BitMagic, the \texttt{vector}, and so forth.

With this in mind, we can reduce the problem of computing the resulting cardinalities of various operations to the problem of computing the size of the intersection:
\begin{itemize}
\item The size of the union is $|A|+|B|-|A\cap B|$.
\item The size of the difference is $|A|-|A\cap B|$.
\item The size of the symmetric difference is $|A|+|B|- 2 |A\cap B|$.
\end{itemize}

We present the time required to compute the size of the intersection in Table~\ref{tab:intersectioncount}. Compared to Table~\ref{tab:intersection} where we present the time necessary to compute the full intersection and measure its size, these new timings can be twice as small or better. Unsurprisingly, it is efficient to avoid materializing the intersection.

The results are otherwise somewhat similar: except for a few datasets, Roaring is best.
However, we observe that BitMagic is much closer in performance to Roaring, even surpassing it slightly in one dataset, than
in Table~\ref{tab:intersection}. As reported in \S~\ref{sec:bitmagic}, the BitMagic library includes optimizations specific to the computation of the cardinality.

In Table~\ref{tab:perfcount}, we also present results for the computation of the sizes of the unions, differences and symmetric differences of integer sets. We can compare these timings with Table~\ref{tab:perf} where we compute the complete results.

 \begin{table*}
\caption{\label{tab:perfcount}Performance comparisons for computing the sizes of the intersections, unions, differences and symmetric differences of integer sets. We report the number of CPU cycles used, divided by the number of input values. For each dataset, the best result is in bold.}
\centering


\subfloat[\label{tab:intersectioncount}Two-by-two intersections: given 200~sets, we compute the sizes of 199~intersections, without materializing the intersections.]{%
\centering\small
\begin{tabular}{c|rrrr>{\columncolor[gray]{0.9}}rrrr}
\toprule
 & bitset& \texttt{vector}  & hash set & BitMagic & Roaring & EWAH & WAH & Concise\\
        \midrule
\CensInc{}                 & \textbf{0.090}    & 4.85              & 13.3              & 0.140             & 0.130             & 0.800             & 0.980             & 1.25              \\
\CensIncsrt{}              & \textbf{0.090}    & 2.08              & 6.76              & 0.120             & 0.110             & 0.190             & 0.260             & 0.340             \\
\Censeighten{}             & 6.31              & 1.45              & 0.460             & 0.090             & \textbf{0.080}    & 3.71              & 9.04              & 12.1              \\
\Censeightensrt{}          & 11.4              & 1.70              & 1.05              & 0.120             & \textbf{0.060}    & 0.440             & 1.02              & 1.27              \\
\Weather{}                 & 0.270             & 4.54              & 17.5              & \textbf{0.210}    & 0.260             & 1.64              & 2.09              & 2.75              \\
\Weathersrt{}              & 0.190             & 1.97              & 7.19              & 0.090             & \textbf{0.060}    & 0.150             & 0.220             & 0.260             \\
\Wikileaks{}               & 11.7              & 2.12              & 12.8              & 1.57              & \textbf{0.870}    & 2.46              & 3.85              & 4.11              \\
\Wikileakssrt{}            & 8.71              & 1.69              & 10.4              & 0.510             & \textbf{0.260}    & 0.540             & 0.760             & 0.830             \\
\bottomrule
\end{tabular}
}\\
\subfloat[\label{tab:unioncount}Two-by-two unions: given 200~sets, we compute the sizes of 199~unions, without materializing the unions.]{%
\centering\small
\begin{tabular}{c|rrrr>{\columncolor[gray]{0.9}}rrrr}
\toprule
 & bitset& \texttt{vector}  & hash set & BitMagic & Roaring & EWAH & WAH & Concise\\
        \midrule
\CensInc{}                 & \textbf{0.070}    & 4.62              & 13.4              & 0.370             & 0.130             & 1.11              & 0.990             & 1.32              \\
\CensIncsrt{}              & \textbf{0.070}    & 1.87              & 6.76              & 0.200             & 0.120             & 0.290             & 0.280             & 0.380             \\
\Censeighten{}             & 7.62              & 1.27              & 0.220             & 1.18              & \textbf{0.100}    & 12.1              & 11.6              & 16.8              \\
\Censeightensrt{}          & 12.6              & 1.57              & 0.840             & 0.520             & \textbf{0.140}    & 1.06              & 1.28              & 1.65              \\
\Weather{}                 & \textbf{0.230}    & 4.36              & 17.6              & 0.570             & 0.270             & 2.50              & 2.10              & 2.92              \\
\Weathersrt{}              & 0.180             & 1.75              & 7.20              & 0.140             & \textbf{0.060}    & 0.230             & 0.240             & 0.300             \\
\Wikileaks{}               & 11.8              & 2.37              & 10.6              & 2.22              & \textbf{1.33}     & 4.08              & 4.46              & 4.89              \\
\Wikileakssrt{}            & 10.5              & 1.57              & 9.40              & 1.16              & \textbf{0.490}    & 1.00              & 1.02              & 1.16              \\
\bottomrule
\end{tabular}
}
\\
\subfloat[\label{tab:differencecount}Two-by-two differences: given 200~sets, we compute  the sizes of 199~differences, without materializing the differences.]{%
\centering\small
\begin{tabular}{c|rrrr>{\columncolor[gray]{0.9}}rrrr}
\toprule
 & bitset& \texttt{vector}  & hash set & BitMagic & Roaring & EWAH & WAH & Concise\\
        \midrule
\CensInc{}                 & \textbf{0.090}    & 5.21              & 47.4              & 0.280             & 0.130             & 0.950             & 1.00              & 1.26              \\
\CensIncsrt{}              & \textbf{0.100}    & 2.28              & 29.5              & 0.190             & 0.120             & 0.230             & 0.280             & 0.350             \\
\Censeighten{}             & 8.26              & 1.63              & 44.6              & 1.60              & \textbf{0.080}    & 8.14              & 10.8              & 14.5              \\
\Censeightensrt{}          & 14.2              & 1.99              & 21.5              & 3.19              & \textbf{0.090}    & 0.720             & 1.04              & 1.28              \\
\Weather{}                 & 0.280             & 5.11              & 61.4              & 0.450             & \textbf{0.270}    & 2.05              & 2.14              & 2.77              \\
\Weathersrt{}              & 0.210             & 2.19              & 32.3              & 0.170             & \textbf{0.060}    & 0.190             & 0.230             & 0.270             \\
\Wikileaks{}               & 13.5              & 2.53              & 40.3              & 5.27              & \textbf{1.09}     & 3.28              & 4.15              & 4.29              \\
\Wikileakssrt{}            & 11.0              & 1.88              & 25.8              & 3.64              & \textbf{0.370}    & 0.790             & 0.950             & 1.03              \\
\bottomrule
\end{tabular}
}
\\
\subfloat[\label{tab:symdiffcount}Two-by-two symmetric differences: given 200~sets, we compute  the sizes of 199~symmetric differences, without materializing the symmetric differences.]{%
\centering\small
\begin{tabular}{c|rrrr>{\columncolor[gray]{0.9}}rrrr}
\toprule
 & bitset& \texttt{vector}  & hash set & BitMagic & Roaring & EWAH & WAH & Concise\\
        \midrule
\CensInc{}                 & \textbf{0.090}    & 4.80              & 13.3              & 0.390             & 0.130             & 1.13              & 0.980             & 1.30              \\
\CensIncsrt{}              & \textbf{0.100}    & 2.10              & 6.77              & 0.220             & 0.120             & 0.310             & 0.280             & 0.380             \\
\Censeighten{}             & 9.79              & 1.46              & 0.470             & 1.11              & \textbf{0.080}    & 12.0              & 11.6              & 16.9              \\
\Censeightensrt{}          & 16.3              & 1.78              & 0.800             & 0.520             & \textbf{0.130}    & 1.07              & 1.27              & 1.66              \\
\Weather{}                 & 0.290             & 4.51              & 17.6              & 0.600             & \textbf{0.270}    & 2.52              & 2.09              & 2.90              \\
\Weathersrt{}              & 0.220             & 1.97              & 7.19              & 0.140             & \textbf{0.060}    & 0.240             & 0.240             & 0.300             \\
\Wikileaks{}               & 14.8              & 2.23              & 11.3              & 2.21              & \textbf{1.29}     & 4.09              & 4.47              & 4.93              \\
\Wikileakssrt{}            & 12.8              & 1.70              & 8.77              & 1.14              & \textbf{0.470}    & 0.990             & 1.02              & 1.16              \\
\bottomrule
\end{tabular}
}
\end{table*}


\subsection{Effect of our Optimizations}


When compiling scalar C code, the compiler and its standard library use advanced (e.g., SIMD) instructions. However, we are interested in the effect of our specific SIMD optimizations (see \S~\ref{sec:bitmanip} and \S~\ref{sec:vecproc}).
Within the CRoaring library, we can deactivate these
optimizations at compile time, falling back on portable C code. The compiler still uses advanced instructions, but without our direct assistance.
In  Table~\ref{tab:ouropti}, we present a comparison, using two
datasets where the optimizations were especially helpful. In some cases, we double the speed, or better. In these datasets, the functions to compute the size (or count) of the intersection, union, difference and symmetric differences were all similarly helped. In other datasets, our optimizations had smaller benefits or no benefit at all %
(though they did no harm).   

We find that these results stress the value of our optimizations for several reasons:
\begin{itemize}
\item As already stated, even when our optimizations are disabled, the compiler still relies on advanced (SIMD) instructions.
\item We are attempting to optimize already efficient code: without our optimizations, CRoaring is already faster than most alternatives in our benchmarks.
\item Our optimizations  focus on only some specific cases (see \S~\ref{sec:vecproc}). The bulk of the code base remains the same when we deactivate our optimizations. For this reason, we should not expect them to help in all or even most benchmarks. For example, none of the operations involving run containers have received specific optimizations.
\end{itemize}



\begin{table}
\caption{\label{tab:ouropti}CPU cycles used per input element before and after our SIMD-based optimizations (\S~\ref{sec:vecproc})}
\centering\small
\begin{tabular}{ll|rrr}
\toprule
 &  & \CensInc{}  & \Weather{} \\
        \midrule
\multirow{2}{*}{\pbox{3.5cm}{2-by-2 
intersections}} & scalar code  & 0.450 & 0.990 \\
 & SIMD code   & 0.220 & 0.380\\\cdashline{2-4}
  & scalar/SIMD   & 2.0 & 2.6\\[5pt]
\multirow{2}{*}{\pbox{3.5cm}{2-by-2 
unions}} & scalar code  & 0.410 & 0.880 \\
 & SIMD code  & 0.270 & 0.570 \\\cdashline{2-4}
   & scalar/SIMD   & 1.5 & 1.5\\[5pt]
\multirow{2}{*}{\pbox{3.5cm}{2-by-2 
difference}} & scalar code  & 0.520 & 1.10 \\
 & SIMD code  & 0.280 & 0.440 \\\cdashline{2-4}
   & scalar/SIMD   & 1.9 & 2.5\\[5pt]
\multirow{2}{*}{\pbox{3.5cm}{2-by-2 
sym.\ diff.}} & scalar code  & 0.540 & 1.06 \\
 & SIMD code  & 0.370 & 0.760 \\\cdashline{2-4}
   & scalar/SIMD   & 1.5 & 1.4\\[5pt]
\multirow{2}{*}{\pbox{3.5cm}{2-by-2 
intersection counts}} & scalar code  & 0.360 & 0.890 \\
 & SIMD code   & 0.130 & 0.260\\\cdashline{2-4}
  & scalar/SIMD   & 2.8 & 3.4\\[5pt]
\multirow{2}{*}{wide union} & scalar code  & 0.060 & 0.250 \\
 & SIMD code  & 0.050 & 0.160 \\\cdashline{2-4}
  & scalar/SIMD   & 1.2 & 1.6\\
\bottomrule
\end{tabular}
\end{table}

\section{Conclusion}

No single approach can be best in all cases, but Roaring offers good all-around performance in our tests as was already reported~\cite{goroaring2017,lemire2016consistently}. Our advanced SIMD-based optimizations further improved already good speeds in several cases.  Other similar data structures (like BitMagic) could make use of our optimizations.

We have identified two relative weaknesses of our implementation that should be the subject of future work:
\begin{itemize}
\item A limitation of our implementation is the lack of advanced optimizations for all operations between a run container and another container type, including another run container. We have determined, by profiling, that these operations  are frequently a significant cost (see Appendix~\ref{appendix:hot}), so they may offer an avenue for further performance gains. In particular, it may prove interesting to attempt to vectorize operations over run containers.
\item When comparing BitMagic and Roaring, two relatively similar designs, we found that BitMagic often has  better random access performance (membership queries), e.g., by up to about  a factor of three. This can be explained by the more compact layout used by Roaring. Careful optimizations might be able to improve Roaring's performance on these queries~\cite{khuong2015array}.
\end{itemize}

Our investigations have focused on  recent x64 microarchitectures. Our CRoaring library can run on a wide range of processors. In the future, we might investigate performance issues on other important architectures such as ARM.

Several features of CRoaring will be reviewed in the future. For example,
CRoaring supports copy-on-write at the container level, meaning that we can copy Roaring bitmaps, in part or in full,  quickly. Given that we have determined that memory allocations and data copies can sometimes be a significant cost (see Appendix~\ref{appendix:hot}), it seems that this might be helpful.  However, the copy-on-write capability adds
implementation complexity.
Moreover, CRoaring is designed to allow memory-file mapping, thus potentially saving extraneous memory allocations when bitmaps are retrieved from external memory. We can potentially combine copy-on-write containers with memory-file mapping.

 Bitmaps can do more than just intersections, differences, symmetric differences and unions, they can support advanced queries such as top-$k$~\cite{SPE:SPE2289}.  Further work should expand on our experimental validation.

The BitMagic library is similar to CRoaring. However, one distinguishing feature of BitMagic is the ability to produce highly compressed serialized outputs for storage on disk. Future work could consider adding this aspect to CRoaring. Moreover, CRoaring could---like BitMagic---use tagged pointers and a customized memory allocator for reduced memory usage and better speed.

One possible application for CRoaring would be as a drop-in replacement for an existing Java library through the Java Native Interface (JNI). Future work could assess the potential gains  and downsides from such a substitution (see Appendix~\ref{appendix:javavsc}).

\section{Acknowledgements}

We thank the author of the BitMagic library (A. Kuznetsov)
for reviewing our manuscript and helping us tune the BitMagic benchmark.


%

\bibliographystyle{wileyj}
\bibliography{croaring}

\FloatBarrier

\appendix

\section{Profiling of Roaring Bitmap Computations}

\label{appendix:hot}
As is typical, given a dataset and an operation, the software spends most of its time in a few functions spanning few lines of code. Using Linux's \texttt{perf} tool, we can identify these hot functions using our test platform and datasets (see \S~\ref{sec:exp}). As part of our publicly available benchmark package, we include a script to automatically run a profiling benchmark.
A detailed analysis would be difficult, as there are too many details to take into account. However, we can at least identify the most expensive operation. We also report the percentage of the total time spent in the corresponding operation. This percentage should be viewed as a rough approximation. When the percentage is relatively low (e.g, 20\%), then it is an indication that there is no clear bottleneck and that many functions contribute to the total running time.
Though we collected detailed data on all operations and all datasets, we only present the results for some key operations, for simplicity (see Table~\ref{tab:hotfunc}). The various benchmarks and datasets are presented in our experimental section (\S~\ref{sec:exp}).

In several instances, the same operation is identified as most significant both with regular code and with  optimizations (as per \S~\ref{sec:bitmanip} and \S~\ref{sec:vecproc}), but the percentage is lower with  optimizations. This reflects the fact that the optimizations were successful at reducing the running time. In cases where the function does not benefit from optimizations, the percentage is mostly unchanged, as expected. It is the case when run containers are involved, as operations over these containers have not benefited from particular optimizations.

In some instances, e.g., for the union between bitmaps in the \Censeighten{}  data set, the \texttt{memcpy} function as well as accompanying memory allocations become a significant cost because we clone many containers. In others cases (e.g., \Censeightensrt{}, \Wikileakssrt{}), memory allocation is significant.

When intersecting bitmaps in the \Censeightensrt{} dataset, much of the time is spent accessing and comparing the high-level 16-bit keys, without access to the containers.

\begin{table}
\caption{\label{tab:hotfunc}Most expensive operations for each benchmark and dataset as determined by software profiling}
\centering\small
\begin{tabular}{cp{0.4\textwidth}p{0.4\textwidth}}
\toprule
2-by-2 intersection & regular code & with optimizations  \\
        \midrule
\CensInc   & array-array (56\%) &  array-array (19\%) \\
\CensIncsrt  &  array-run (42\%) &  array-run (45\%)   \\
\Censeighten   & galloping array-array (38\%)  & galloping array-array (39\%)  \\
\Censeightensrt   & high-level  (26\%) & high-level  (26\%) \\
\Weather  & array-array (72\%) &  array-array (35\%) \\
\Weathersrt  &  array-run (30\%) &  array-run (31\%)   \\
\Wikileaks  & run-run (38\%) &  run-run (38\%)  \\
\Wikileakssrt    & run-run (19\%) &  run-run (20\%)  \\
\bottomrule
\toprule
2-by-2 union & regular code & with optimizations \\
         \midrule
\CensInc   & array-array (47\%) & array-array (20\%) \\
\CensIncsrt  &  array-run (44\%) &  array-run (46\%)   \\
\Censeighten   & \texttt{memcpy} (20\%)  & \texttt{memcpy} (19\%)  \\
\Censeightensrt   & memory allocation (20\%) &  memory allocation (20\%)  \\
\Weather   & array-array (49\%) & bitset$\to$array conversion (23\%) \\
\Weathersrt  &  array-run (26\%) &  array-run (25\%)   \\
\Wikileaks  & run-run (20\%) &  run-run (20\%)  \\
\Wikileakssrt    & memory allocation (15\%) &  memory allocation (13\%)  \\
\bottomrule
\toprule
2-by-2 diff. & regular code & with optimizations \\
        \midrule
\CensInc   & array-array (52\%) & bitset$\to$array conversion (18\%) \\
\CensIncsrt  &  array-run (25\%) &  array-run (27\%)   \\
\Censeighten   & array-array (19\%)  & \texttt{memcpy} (18\%)  \\
\Censeightensrt   &  memory allocation (20\%) &  memory allocation (20\%)  \\
\Weather  & array-array (26\%) & array-array (25\%) \\
\Weathersrt  &  run-run (22\%) &  run-run (22\%)   \\
\Wikileaks  & run-run (24\%) &  run-run (22\%)  \\
\Wikileakssrt    &  memory allocation (14\%) &  memory allocation (16\%)  \\
\bottomrule
\toprule
2-by-2 sym.\ diff. & regular code & with optimizations \\
        \midrule
\CensInc   & array-array (52\%) & bitset-array  (24\%) \\
\CensIncsrt  &  array-array (21\%) & run-run (18\%) \\
\Censeighten   & \texttt{memcpy} (21\%)  & \texttt{memcpy} (21\%)  \\
\Censeightensrt   &  memory allocation (20\%) &  memory allocation (20\%)  \\
\Weather    & array-array (72\%) & array-array (40\%) \\
\Weathersrt    &  run-run (22\%) &  run-run (21\%)   \\
\Wikileaks  & run-run (14\%) &  run-run (14\%)  \\
\Wikileakssrt    &  memory allocation (15\%) &  memory allocation (15\%)  \\
\bottomrule
\toprule
single-union & regular code & with optimizations \\
        \midrule
\CensInc   & bitset-array (56\%) & bitset-array  (52\%) \\
\CensIncsrt  & run-bitset (62\%) &  run-bitset (70\%) \\
\Censeighten   & bitset-array (81\%) & bitset-array  (36\%) \\
\Censeightensrt   & bitset$\to$array conversion (61\%)  & bitset$\to$array conversion (63\%) \\
\Weather   & bitset-array (80\%) & bitset-array  (78\%) \\
\Weathersrt  & run-bitset (66\%) &  run-bitset (68\%) \\
\Wikileaks  &  run-bitset (64\%) &  run-bitset (66\%) \\
\Wikileakssrt   & run-bitset (25\%),  bitset$\to$array  (25\%)  &  run-bitset (26\%),  bitset$\to$array  (26\%)  \\
\bottomrule
\toprule
intersection-count & regular code & with optimizations \\
        \midrule
\CensInc   & array-array (71\%) &  array-array (29\%) \\
\CensIncsrt  &  array-run (55\%) &  array-run (62\%)   \\
\Censeighten   & galloping array-array (54\%)  & galloping array-array (62\%)  \\
\Censeightensrt   & high-level  (46\%) & high-level  (47\%) \\
\Weather   & array-array (81\%) &  array-array (45\%) \\
\Weathersrt  & run-run (42\%) &   run-run (46\%) \\
\Wikileaks  &  run-run (66\%) &   run-run (66\%) \\
\Wikileakssrt   &  run-run (47\%) &   run-run (52\%) \\
\bottomrule
\end{tabular}
\end{table}

\section{Large In-Memory Dataset}

\label{appendix:big}

To validate our results on  a large dataset, we generated 100~arrays of 10~million distinct integers in the range $[0,10^9)$ using the
ClusterData distribution from  Anh and Moffat~\cite{Anh:2010:ICU:1712666.1712668}.
This distribution leaves relatively small gaps between successive integers, with occasional large gaps. As with our main experiments, all our source code and scripts are available online. We ran these experiments using the same hardware and software as in \S~\ref{sec:exp}---using a separate script that is included with our benchmarking software.

In total, we generated one billion integers; thus our benchmarks could use gigabytes of memory (RAM).
We excluded the hash-set implementation   (\texttt{std::unordered_set}) from these tests because it failed to complete all of them without exceeding the memory available on our benchmarking machine.


We present our comparison in Table~\ref{tab:bigresult}. Compared with our main experiments (on smaller datasets), the most obvious difference is that the membership queries are significantly worse for the run-length encoded schemes (EWAH, WAH and Concise), which is not surprising: that membership queries are resolved by scanning the data from the beginning of the set each time in these cases. To a lesser extent, Roaring also suffers from a worse performance for membership queries when compared to the bitset or to BitMagic.

We compare Roaring without our manual SIMD-based optimizations with the scalar code in Table~\ref{tab:ouroptibig}. Our SIMD-based optimizations are worthwhile, sometimes multiplying the performance by a factor of five.

\begin{table*}
\caption{\label{tab:bigresult}Memory usage in bits per value as well as timings in cycles per value for
various operations for the large ClusterData in-memory dataset. For each metric, the
best reported result is in bold. For the \texttt{vector}, sequential access is practically free. }
\centering\small
\begin{tabular}{c|rrr>{\columncolor[gray]{0.9}}rrrr}
\toprule
               & bitset  & \texttt{vector}   & BitMagic & Roaring          & EWAH          & WAH  & Concise\\ \midrule
  bits/value   & 100     &              32.0 &   38.5   & \textbf{13.5}    &   33.4      & 33.7 & 22.6   \\
seq. access     & 28.9 & --- & 46.3 & \textbf{6.87} &25.5 & 27.3 & 27.6 \\
membership   & \textbf{4.92} & 1910 & 82.0 & 381 &\num{19900000} & \num{58600000} & \num{48000000} \\
intersections   & 25.4 & 11.2 & 9.11 & \textbf{2.58} &18.0 & 24.5 & 27.8 \\
unions   & 25.4 & 14.5 & 9.77 & \textbf{4.97} &33.5 & 34.9 & 37.5 \\
differences   & 25.3 & 14.4 & 8.05 & \textbf{2.48} &26.2 & 31.4 & 33.0 \\
sym. differences   & 25.3 & 16.6 & 9.91 & \textbf{4.09} &32.6 & 34.8 & 37.8 \\
wide unions   & 5.01 & 457 & 3.92 & \textbf{3.87} &53.0 & 68.7 & 69.5 \\
intersection counts   & 2.27 & 8.23 &   5.16 & \textbf{1.37} &14.0 & 16.2 & 18.9 \\
union counts   & 2.14 & 8.92 &  8.89 & \textbf{1.35} &17.7 & 16.5 & 19.3 \\
diff. counts  & 2.28 & 9.27 & 7.24 & \textbf{1.37} &16.5 & 16.8 & 18.9 \\
sym. diff. counts  & 2.27 & 8.29 & 9.17 & \textbf{1.35} & 17.8 & 16.6 & 19.2 \\
\bottomrule
\end{tabular}
\end{table*}

\begin{table}
\caption{\label{tab:ouroptibig}CPU cycles used per input element before and after our SIMD-based optimizations (\S~\ref{sec:vecproc})}
\centering\small
\begin{tabular}{ll|rr}
\toprule
 &  & ClusterData  \\
        \midrule
\multirow{2}{*}{\pbox{3.5cm}{2-by-2 
intersections}} & scalar code  & 8.11 \\
 & SIMD code   & 2.58\\\cdashline{2-3}
  & scalar/SIMD   & 3.1\\[5pt]
\multirow{2}{*}{\pbox{3.5cm}{2-by-2 
unions}} & scalar code  &  9.34  \\
 & SIMD code  & 4.97  \\\cdashline{2-3}
   & scalar/SIMD   & 1.9 \\[5pt]
\multirow{2}{*}{\pbox{3.5cm}{2-by-2 
difference}} & scalar code  & 8.33  \\
 & SIMD code  & 2.48  \\\cdashline{2-3}
   & scalar/SIMD   & 3.4 \\[5pt]
\multirow{2}{*}{\pbox{3.5cm}{2-by-2 
sym.\ diff.}} & scalar code  & 9.19 \\
 & SIMD code  &4.09 \\\cdashline{2-3}
   & scalar/SIMD   & 2.2 \\[5pt]
\multirow{2}{*}{\pbox{3.5cm}{2-by-2 
intersection counts}} & scalar code  &6.82  \\
 & SIMD code   & 1.37\\\cdashline{2-3}
  & scalar/SIMD   & 5.0  \\[5pt]
\multirow{2}{*}{wide union} & scalar code  & 4.79 \\
 & SIMD code  & 3.87 \\\cdashline{2-3}
  & scalar/SIMD   & 1.2 \\
\bottomrule
\end{tabular}
\end{table}

\section{Java Versus C}

\label{appendix:javavsc}

In earlier work, we
used Java to
benchmark Roaring against several other set data structures, including 32-bit EWAH~\cite{lemire2016consistently}.  There are many differences between Java and C that make comparisons difficult with our current results in C. For example, Java uses a garbage collector whereas C does not. Nevertheless, we can run the %
SIMD-optimized C
 and Java versions side-by-side on the same machine and report the results. We make our dual-language benchmark freely available to ensure reproducibility (\url{https://github.com/lemire/java_vs_c_bitmap_benchmark}). For simplicity we focus on a single dataset (\Censeighten{}).
On our testing hardware (\S~\ref{sec:hardware}), we use Oracle's JDK 1.8.0_101 for  Linux (x64),  the JavaEWAH library (version~1.0.6) and the RoaringBitmap library (version~0.6.51). We use Oracle's JMH benchmarking system to record the average wall-clock time in Java after many warm-up trials. Our configuration differs in many ways from our earlier work, so results are not directly comparable~\cite{lemire2016consistently}.

 Because the Java and C benchmarks are written using entirely different software, and because the Java timings rely on an extensive framework (JMH), we cannot be certain that the Java-to-C comparison is entirely fair, especially for small timings. For example, the Java benchmark may not account for the memory allocation in the same manner as the C timings.
 To help improve fairness, we added a few ``warm-up'' passes in C/C++, just like the warm-up trials generated in Java by JMH: it helped performance slightly. The small beneficial effect is probably related to caching and branch prediction.

 We report timings (in microseconds) and as well %
give the speed ratio between 32-bit EWAH and Roaring
in Table~\ref{tab:perfjava}.
 The relative benefits of Roaring over EWAH in these tests are sometimes more important in C/C++ and  sometimes more important in Java. The Java code is slower by up to a factor of two, but the difference between Java and C can also  be negligible.
The performance differences between Roaring and EWAH in these tests dominate the differences that are due to the programming languages and implementations. That is, the Java version of Roaring is several times faster than the C++ version of EWAH.

 \begin{table*}
\caption{\label{tab:perfjava}Performance comparisons for intersections between Java and C/C++ implementations. We report timings in microseconds for executing the whole benchmark (e.g., compute all pairwise intersections). A ratio greater than 1 indicates that Roaring is faster.}
\centering\small
\begin{tabular}{c|rrr|rrr|rr}
\toprule
 & \multicolumn{3}{c}{C/C++} & \multicolumn{3}{c}{Java}  & \multicolumn{2}{c}{Java/C ratio} \\
 & EWAH & Roaring & ratio & EWAH & Roaring & ratio & EWAH & Roaring\\
        \midrule
2-by-2 intersections  & \num{1900} & 42 & 45 & \num{2480} & 43 & 58 & 1.3 & 1.0 \\
2-by-2 unions         & \num{12800} & 605      & 21 & \num{28500} & 750 & 38& 2.2  & 1.2 \\
wide union            & \num{77600} & 677         & 114 & \num{107000} & 1240 & 86 & 1.4 & 1.8 \\

\bottomrule
\end{tabular}
\end{table*}

\end{document}